\documentclass[prl,superscriptaddress,showpacs,twocolumn,nofootinbib,floatfix,amsmath,amssymb]{revtex4-1}
\usepackage{amsmath, amsthm, amstext, amsfonts, hyperref, graphicx, braket}

\theoremstyle{definition}

\renewcommand{\d}{\mathrm{d}}
\newcommand{\ii}{\mathrm{i}}

\allowdisplaybreaks
\usepackage{color}

\begin{document}


\title{Quantum gates via relativistic remote control}


\author{Eduardo Mart\'in-Mart\'inez}
\affiliation{Institute for Quantum Computing, University of Waterloo,  Waterloo, Ontario, N2L 3G1, Canada}
\affiliation{Dept. Applied Math., University of Waterloo, Ontario, N2L 3G1, Canada}
\affiliation{Perimeter Institute for Theoretical Physics, Waterloo, Ontario N2L 2Y5, Canada}
\author{Chris Sutherland}
\affiliation{Institute for Quantum Computing, University of Waterloo,  Waterloo, Ontario, N2L 3G1, Canada}


\begin{abstract}
We harness relativistic effects to  gain quantum control on a stationary qubit in an optical cavity by controlling the non-inertial motion of a different probe atom. Furthermore, we show that by considering relativistic trajectories of the probe, we enhance the efficiency of the quantum control. We explore the possible use of these relativistic techniques to build 1-qubit quantum gates.
\end{abstract}

\maketitle



{\it Introduction.-}
The study of the interface between quantum mechanics, field theory and general relativity has led to results where, in principle, relativistic features can be used to gain advantage over non-relativistic settings in the processing of quantum information \cite{MigC,aasen,bruschi,Prdvette}.

To implement quantum gates, or even quantum simulators, we need to very accurately control the degrees of freedom we use as qubits as well as the dynamics of the quantum mechanical systems that contain them. Such degree of control has been achieved, for instance,  in NMR devices \cite{nmr} which  have been largely employed to implement quantum computing algorithms on nuclear spins. 

 In these devices, electrical currents are used to generate magnetic fields that ultimately influence the state of the nuclear spin qubit. The microscopic mechanism of how the accelerated charges interact with the nuclear spin is commonly simplified by treating the field classically. However, it is not unreasonable to think that detailed study of the interaction of the moving charges with the qubit degrees of freedom --mediated by a fully quantum EM field-- may enhance our ability to control the qubit. Moreover, treating this setting in a relativistic framework may allow us to see how (or if) relativistic effects can actually improve our capacity  to control the qubit beyond what classical models predict.
 
 From the fundamental high-energy physics point of view this analysis may prove interesting in the following way: We will show that the relativistic motion of a probe induces high-energy relativistic effects that can be used to control a logical qubit stored in a stationary atom. Hence,  this suggests a connection between high energy physics and quantum optics and information. For instance, based on these results, one could think of using charged beams generated by particle colliders to control the state of atomic qubits, and maybe recast some of the problems of measurement of the outcome of particle colliders in terms of quantum informational variables. As we will highlight, the phenomena described in this paper already manifests at the scales of energies present in the LHC.


It is already known that non-inertial motion can be used to implement universal single qubit gates on atomic systems \cite{aasen} and Gaussian two-qubit gates on cavity field modes \cite{bruschi}. In more detail, \cite{aasen} showed that control over the acceleration of atoms can be used to perform quantum gates as a direct consequence of relativistic quantum effects.  However, these schemes  require control over both the internal degrees of freedom of an atom and over the non-inertial motion of its center-of-mass, which may prove challenging in a practical experimental setting. For instance, the force that accelerates the atom may also induce transitions between the energy levels that constitute the logical qubits.

 In this paper we explore how  controlling the trajectory of an accelerated atom (the probe atom), allows us to garner control over a different atomic qubit (the target qubit) that sits stationary inside an optical cavity.
 Namely, we will show that it is indeed possible to perform arbitrary rotations on the Bloch sphere of the state of the target qubit with only a small decoherence effect.  We obtain such effects already in the simplified case of uniformly accelerated trajectories of the probe atom,  even in the relatively simple scenario where we consider only  atoms (one probe and one target) coupled through the interaction with the quantum field.
 
  Furthermore, we show that when the probe is allowed to attain high speeds,  relativistic effects start to influence the target atom. Remarkably, and maybe against intuition \cite{matsako,matsako2}, these effects allow for better control of the target qubit. We will quantitatively show how we can effectively get larger controlled Bloch sphere rotations when the probe's motion is relativistic as opposed to non-relativistic.
 
\begin{figure}
\includegraphics[width=0.15\textwidth]{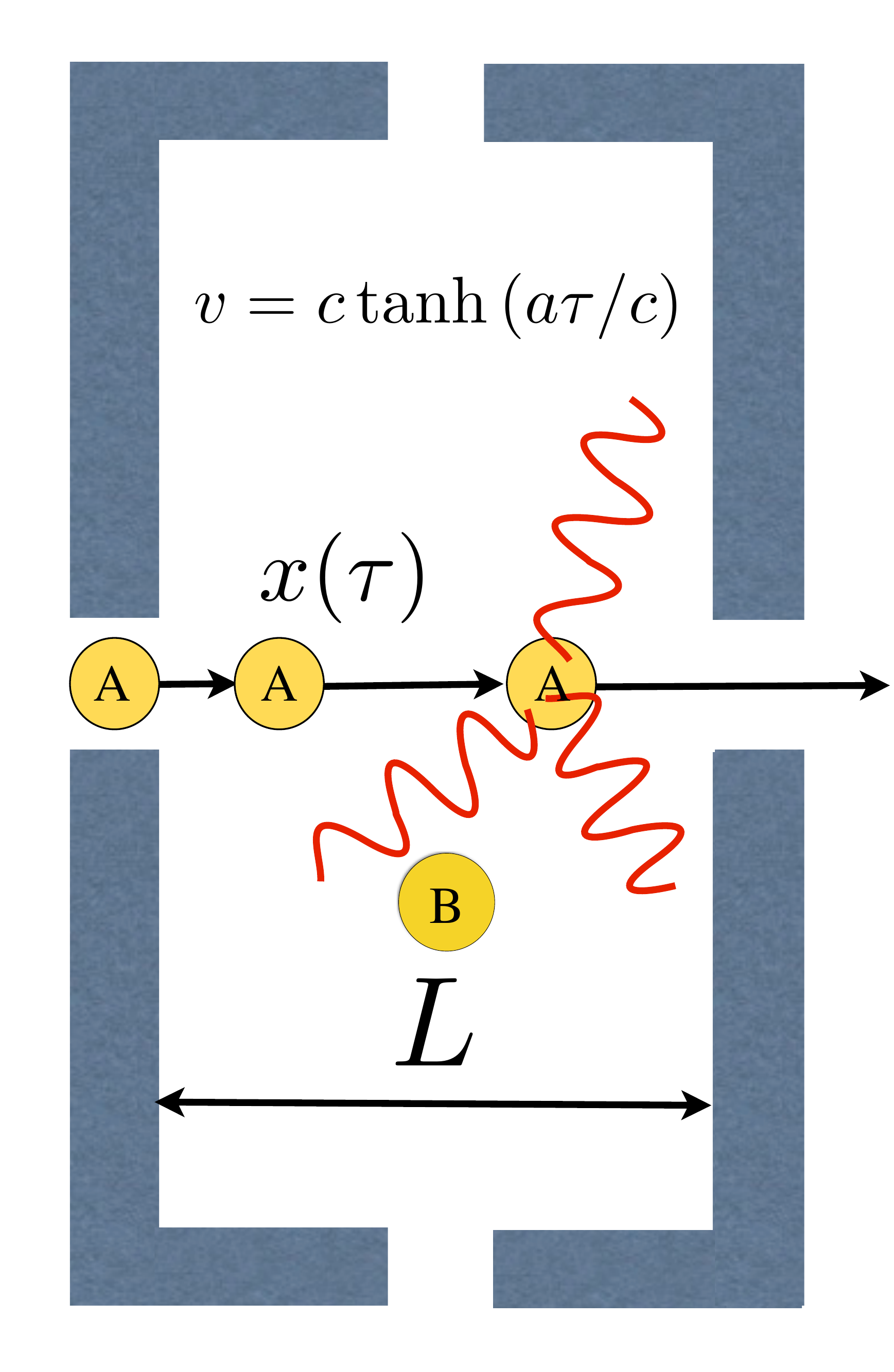}
\caption{The probe atom (A) is shot through the cavity and the target atom (B) is stationary at $x=L/2$, they interact only via the field. We  control  the probe's trajectory, and this gives us control over the target qubit. The probe's worldline is given by $t(\tau)=a^{-1}\sinh a\tau$, $x(\tau)=a^{-1}(\cosh a\tau -1)$ }
\label{cavity}
\end{figure}

{\it Setup.-} We will consider a target atom at rest inside a stationary optical cavity of purely reflective walls as illustrated in Fig. \ref{cavity}. The probe atom will fly through the cavity describing a constantly accelerated motion. Both atoms couple locally (along their respective worldlines) to the quantum field inside the cavity.

We will use the Unruh-DeWitt Hamiltonian  \cite{dewitt} to model the light-matter interaction. This model, often used to model relativistic particle detectors \cite{Crispino}, is identical to the Jaynes-Cummings model of light-matter interaction \cite{ScullyBook} but without taking the single mode approximation nor the rotating wave approximation. Although simple, the model captures all the features of the light-matter interaction when no orbital angular momentum exchange transitions are considered \cite{eduardodewitt,Alvaro}.

The Hamiltonian for a single detector will be of the form $H=H_{0}^{(d)}+H_{0}^{(f)}+H_{I}$, where $H_{0}^{(d)}$ and $H_{0}^{(f)}$ are the detector and field free Hamiltonians. The interaction Hamiltonian $H_{I}$ is of the form $H_{I}=\lambda\xi(\tau)\mu(\tau)\Phi[x(\tau)]$ where $\lambda \xi(\tau)$ is a time dependent coupling strength controlling the interaction time, $\mu(\tau)=(\sigma^+ e^{-\ii\Omega\tau }+\text{H.c.})$ is the monopole moment operator (in the interaction picture) where $\Omega$ is the energy gap between the two levels of the atom, $x(\tau)$ is the worldline of the atom parametrized in terms of its proper time and $\Phi[x(\tau)]$ is the field operator which we expand in terms of stationary wave modes. Throughout the paper we will use natural units $c=\hbar=1$ and we will take the scale $\Omega$ as our reference. How our results translate into  dimensionful units is explained in the section `Experimental feasibility' below. 
  
Since there are two atoms with different states of motion (thus different proper reference frames) we need to choose with respect to what time parameter we want the full Hamiltonian to generate evolution. We choose the proper time of the stationary atom; consequently there is a redshift factor in front of the accelerated atom  term of $H_I$. This is a somewhat subtle point which is discussed in-depth in \cite{brown}. Taking all this into account we finally obtain $ H_I(t)=\frac{\d \tau}{\d t} H^{(\text{A})}_{I}[\tau(t)]+H^{(\text{B})}_{I}(t)$, where the individual $H^{(\text{d})}_{I}$ are given by the single detector interaction Hamiltonian shown above and $t$ is the cavity rest frame time.

We initially prepare the quantum field in the cavity such that one of the field modes is in a coherent state of complex amplitude $\alpha$, and the rest of the modes are lowly-populated. Preparing a near-resonant coherent state reduces the amount of entanglement acquired between the atoms and the field. This in turn helps screen out the mixedness effects on the target qubit produced by the `Unruh noise' generated by the probe's relativistic motion \cite{aasen,Takagi}. Thus the initial atoms-field density matrix can be written as $
\rho_{0}=\rho_{A,0}\otimes\rho_{B,0}\otimes\ket{\alpha_{\omega_{1}}}\bra{\alpha_{\omega_{1}}}\otimes_{\omega_n\neq\omega_{1}}\ket{0_{\omega_n}}\bra{0_{\omega_n}}$. Notice that, since we are in a cavity, the frequencies $\omega_n=n\pi/L$ form a discrete set.

When the probe enters the cavity, it  becomes coupled to the field. We take a perturbative approach (valid for small couplings and short times) to analyze the system dynamics. The time evolution under this Hamiltonian from a time $t=0$ to time $t=T$ is given by
\begin{align}
 \nonumber U(T,0)=&\openone-\ii\int_{0}^{T}\!\!\!\!\d t_{1} H_{I}(t_{1})-\!\int_{0}^{T}\!\!\!\!\d t_{1}\!\!\int_{0}^{t_1}\!\!\!\!\d t_{2}H_{I}(t_1)H_{I}(t_2),
\end{align}
plus terms $\mathcal{O}(\lambda^3)$, where the notation $\mathcal{O}(\lambda^n)$  refers to powers of the coupling strengths of both the probe-field $\lambda_A$ and target-field $\lambda_B$, so that $\lambda_A\lambda_B$ is an $\mathcal{O}(\lambda^2)$ term. The density matrix after a time $T$ will be given by the perturbative expansion $\rho_{T}=\rho_0+\rho^{(1)}_{T}+\rho^{(2)}_{T}+\mathcal{O}(\lambda^3)$ where
\begin{align}
&\rho_{T}^{(1)}=U^{(1)}\rho_{0}+\rho_{0}{U^{(1)}}^\dagger,\\
&\label{secondg}\rho_{T}^{(2)}=U^{(1)}\rho_{0}{U^{(1)}}^\dagger+U^{(2)}\rho_{0}+\rho_{0}{U^{(2)}}^{\dagger}.
\end{align}

Recall that we are interested in the target's final state, and so we will trace out the field modes as well as the probe's state to obtain: $\rho_{T,B}=\text{Tr}_{A}(\text{Tr}_{f}(\rho_{T}))$. We will compare this to the target's initial density matrix, and quantitatively assess our ability to control the target qubit by controlling the probe's motion.

{\it Performing 1-qubit rotations.-} In \cite{aasen}, one-qubit gates were obtained through the non-inertial motion of the atom which supported the logical qubit. Arbitrary rotations on the Bloch sphere were achieved introducing no decoherence to leading order in perturbation theory. The price to pay is that logical quantum operations are performed on the qubit whose non-inertial trajectory had to be controlled. As opposed to \cite{aasen} we use the motion of a \textit{different} probe atom to gain control over the target qubit, physically supported on a different atom which rests in the cavity. Hence, we are not required to keep under control both the trajectory and the internal state of one atom simultaneously. While advantageous in this sense, there is a trade-off on the quality of the quantum gates that we could implement with this setting. As the `remote control' appears as a second order effect, it is impossible to perform a 100\% clean Bloch sphere rotation via this mechanism and, unavoidably, some mixedness will be introduced in the target state. In contrast, in \cite{aasen} the dynamics were fully unitary to leading order in perturbation theory. However, we will show that the mixedness introduced in the stationary qubit is always small as compared to the magnitude of the rotations that we can obtain on the target's Bloch sphere vector. Moreover, we will show that it is indeed advantageous to consider regimes where the probe's trajectory is relativistic in order to more efficiently manipulate the target's qubit. 

First order contributions to the target's time evolution cannot be influenced by the interaction of the field and the probe: At first order in perturbation theory we will only have contributions to the target dynamics which are proportional to $\lambda_B$, and thus these effects are only dependent on the initial state of the field and the target. The leading order contributions to the remote control of the target have to be proportional to $\lambda_A\lambda_B$.

This said, we still need to consider these $\mathcal{O}({\lambda_B})$ contributions to target dynamics. To characterize them, it is convenient to make the following definition:
\begin{align}\label{firstI}
 I_{\pm, j}^{(d)}(T)=\!\!\int_{0}^{T}\!\!\!\!\!\d t\,\frac{ \sin [k_{j}x_{d}(t)]}{\sqrt{\omega_{j}L}}\xi_{d}[\tau_{d}(t)]e^{\ii[\pm\Omega_{d}\tau_{d}(t)+\omega_{j}t]}, 
\end{align}
 where we will absorb the redshift factor $\frac{d\tau}{dt}$  into the probe's switching function  $\xi_{a} (t)$. $U^{(1)}$ then yields 
\begin{align}
\nonumber U^{(1)}= -\ii\sum_{j}&\Big[\lambda_{A}(I_{+,j}^{(A)}a_{j}^{\dagger}\sigma^{+}_{A}+I_{-,j}^{(A)}a_{j}^{\dagger}\sigma^{-}_{A}+\text{H.c.})\\[-3mm]
&  +\lambda_{B}(I_{+,j}^{(B)}a_{j}^{\dagger}\sigma^{+}_{B}+I_{-,j}^{(B)}a_{j}^{\dagger}\sigma^{-}_{B}+\text{H.c.})\Big].
\end{align}
Hence, the first order contribution to $\rho_{T,B}$  is given by
\begin{align}
&  \rho^{(1)}_{T,B}=\text{Tr}_{A,f}(U^{(1)}\rho_{0})+\text{Tr}_{A,f}(\rho_{0}U^{(1)\dagger})=-i\lambda_{B}[\alpha^{*}\\
 &\nonumber\!\!\times\! (I_{+,1}^{(B)}\sigma^{+}_{B}+I_{-,1}^{(B)}\sigma^{-}_{B}) \!+\! \alpha(I_{-,1}^{(B)*}\sigma^{+}_{B}+I_{+,1}^{(B)*}\sigma^{-}_{B})]
\rho_{B,0}\!+\!\text{H.c.},
\end{align}
where, as expected, the $\lambda_A$ terms disappear. 

Note that $\rho^{(1)}_{T,B}$ can actually be expressed as an infinitesimal rotation on the Bloch sphere as shown in \cite{aasen}. This is useful if one is only interested in the moving atom's relativistic effects on the qubit it supports. Since in this paper we are considering the target qubit to be stationary and we want to manipulate its state by controlling an atomic probe's trajectory, we must go to second order in perturbation theory where the `remote-control' terms  (proportional to $\lambda_{A}\lambda_{B}$) naturally arise, i.e. it is necessary to compute $\rho^{(2)}_{T,B}$  from \eqref{secondg}. 
 
 This calculation is algebraically straightforward, but it is rather lengthy and involves 110 non-trivial contribution terms (for pedagogical reasons we include those technically uncomplicated calculations in the appendix).  The reason for the additional complexity can be seen by looking more closely at $U^{(2)}$: There are four terms of the form $\mu_{d}(\tau_{d}(t_{1}))\Phi(x_{d}(\tau_{d}(t_{1})))\mu_{d}(\tau_{d}(t_{2}))\Phi(x_{d}(\tau_{d}(t_{2})))$ acting on $\rho_{0}$ which will all yield many contributions to the target qubit's final state. To give a flavour of the form of the amplitudes multiplying such terms, we use the following convenient short hand notation:
\begin{align} \label{secondJ} &  J_{\pm,\pm,j}^{\mu,\nu}(T)=\frac{1}{\omega_{j}L}\!\int_{0}^{T}\!\!\!\!\d t_{1}\!\int_{0}^{t_1}\!\!\!\d t_{2}\,\xi_{\mu}(\tau_{\mu}(t_{1}))\xi_{\nu}(\tau_{\nu}(t_{2}))\\
 \nonumber & e^{\ii[\Omega_{\mu}\tau_{\mu}(t_{1})\pm\Omega_{\nu}\tau_{\nu}(t_{2})+\omega_{j}(t_{1}\pm t_{2})]}\sin x_{\mu}(\tau_{\mu}(t_{1}))\sin x_{\nu}(\tau_{\nu}(t_2)),
\end{align}
where the labels $\mu,\nu$ can take the values $A,B$. The contributions of $U^{(1)}\rho_0 {U^{(1)}}^\dagger$ are going to be multiplied by the product of two coefficients of the form \eqref{firstI}. On the other hand, the contributions of $U^{(2)}\rho_0 $ and its H.c. will appear multiplied by coefficients of the form \eqref{secondJ}.

\begin{figure}[h] 
\includegraphics[width=0.45\textwidth]{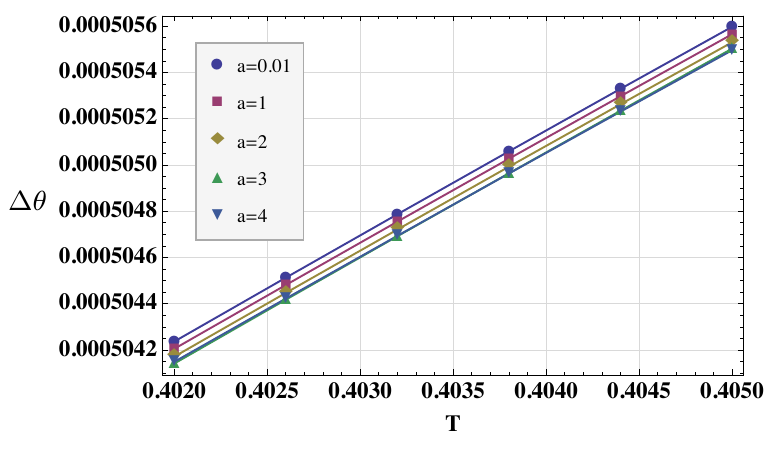}\\
\includegraphics[width=0.45\textwidth]{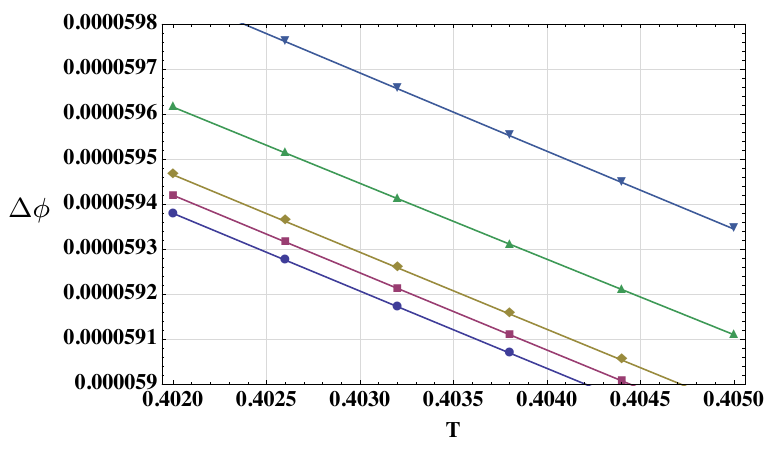}
  \caption{ (Top to bottom) {\bf a)} Change in the target qubit's $\theta$ coordinate for various $a$ as a function of $T$.   {\bf b)} Change in the target qubit's $\phi$ coordinate for various $a$ as a function of $T$. We can obtain independent Bloch sphere rotations by varying specifically chosen values of $a$ and $T$ for the probe atom. Also, variances of the flying time can be used to quickly adjust the magnitude of the rotations, whereas changing the probe's acceleration can be used to fine-tune it.}
  \label{fig:results}
\end{figure}

{\it Results.-}  For simplicity, we choose the initial state of the probe atom to be pure with real amplitudes $\ket{\psi_A}=p\ket{g}+\sqrt{1-p^2}\ket{e}$. We then control the trajectory of the probe through the cavity containing the target qubit prepared in an arbitrary state.  We show in Fig \ref{fig:results} that the indirect interaction between the probe and the target is powerful enough to produce small independent Bloch sphere rotations of the target qubit. These rotations' amplitude and direction are controlled via controlling the parameters of the non-inertial motion of the probe. We can very quickly vary the amount of rotation both in the azimuthal $(\phi)$ and the polar $(\theta)$ direction by controlling the flying time $T$ of the probe atom. This flying time can be controlled independently of $a$ by controlling the atomic probe initial speed before entering the cavity, or alternatively by controlling the cavity length.  In the simulations we run and the plots we plot we used the first method. However, alternatively, the cavity crossing time of the probe (starting at rest) for a constantly accelerated probe is given by $T(L)=a^{-1} \text{arccosh}(aL + 1)$. Notice that this second method consisting of varying the length of the cavity to control $T$ also modifies the mode structure of the field through a shift of the field temporal and spatial frequencies proportional to $1/L$. This implies that if one wants to control the flying time through the length of the cavity one necessarily has to be careful with this subtlety.

 The magnitude of the rotations is more influenced by the probe's flying time than by its acceleration, therefore by controlling its acceleration we can fine tune the qubit rotations. The amount of rotation along the $\theta$ and $\phi$ directions can be  separately controlled by  independently varying the acceleration and flying time of the probe. Notice that although Fig. \ref{fig:results} shows that as $T$ increases, $\Delta\phi$ decreases, and as $a$ increases, $\Delta\theta$ decreases, this is only true in the low $T$ regimes. As we will show below, when the probe's flying time is large enough, relativistic trajectories $aT>1$  perform better than non-relativistic ones. The dependence of $\Delta\theta$ and $\Delta\phi$ with $a$ and $T$ is non-monotonic as we cross over the boundary of the relativistic regime.

\begin{figure*}[t] \begin{tabular}{cccc}
\!\includegraphics[width=0.25\textwidth]{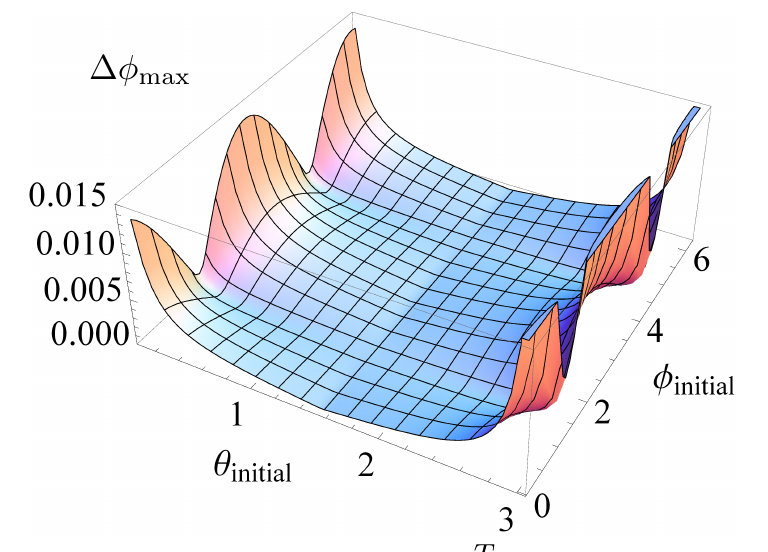}&
\includegraphics[width=0.25\textwidth]{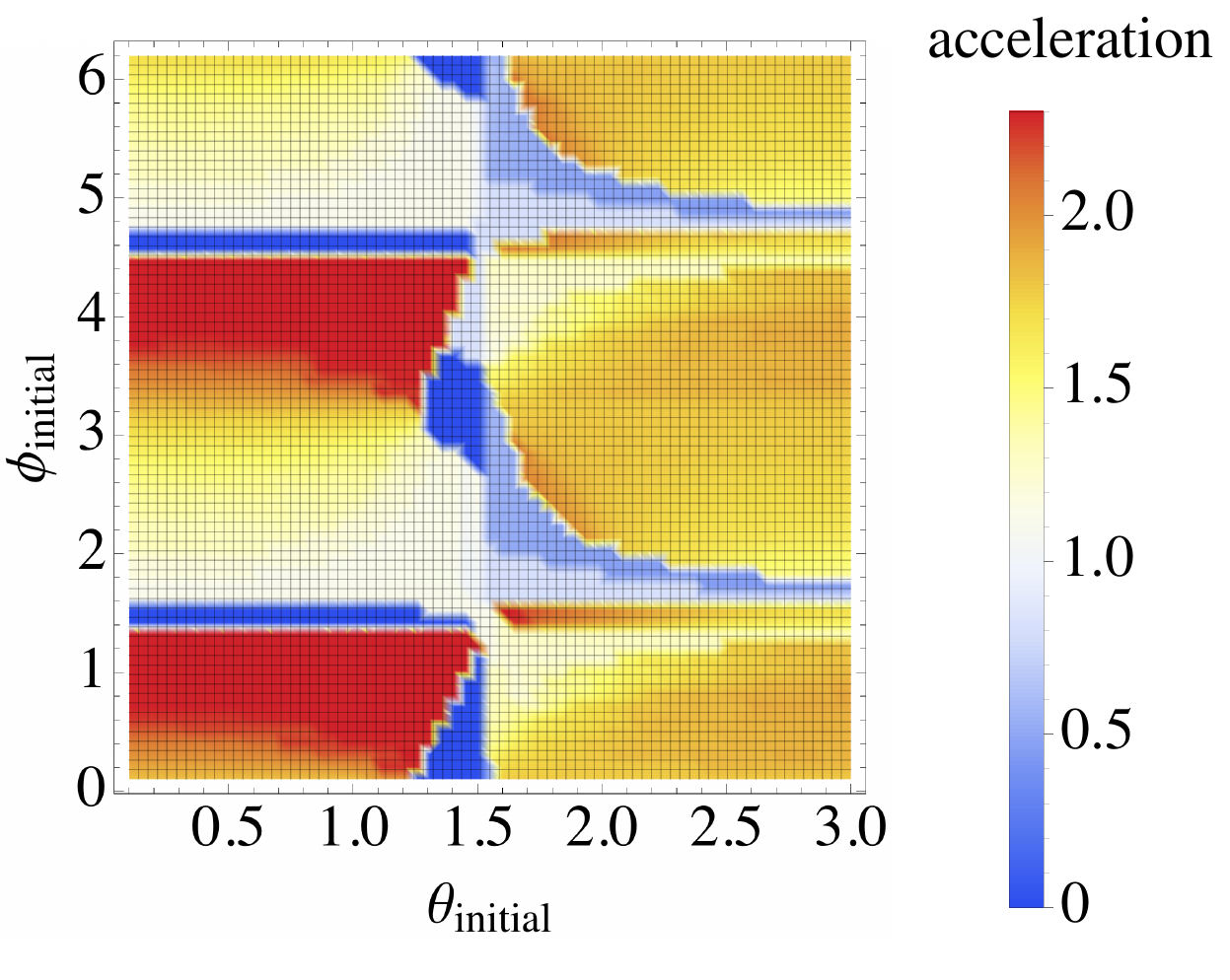}&
\!\!\includegraphics[width=0.25\textwidth]{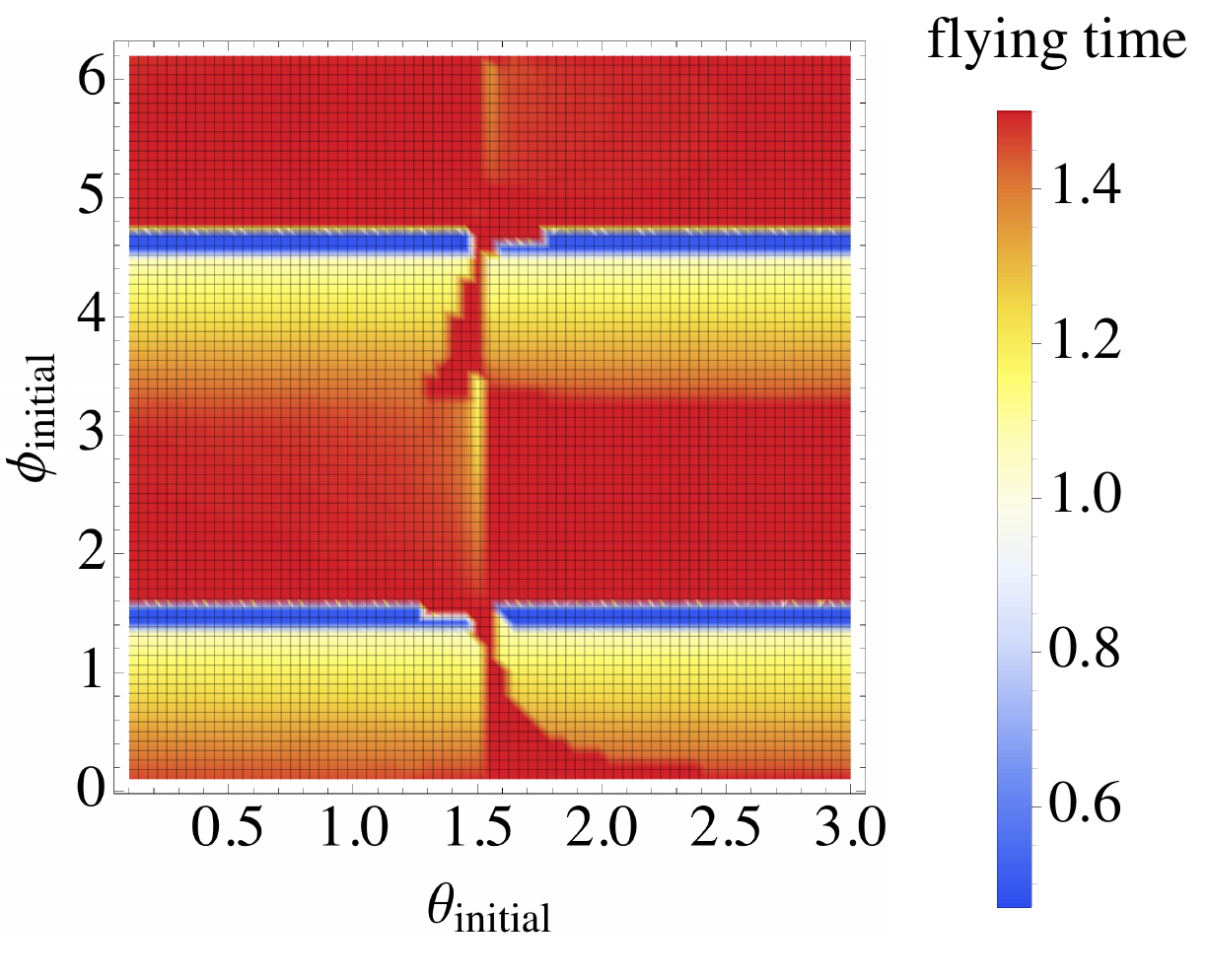}&
\!\!\includegraphics[width=0.25\textwidth]{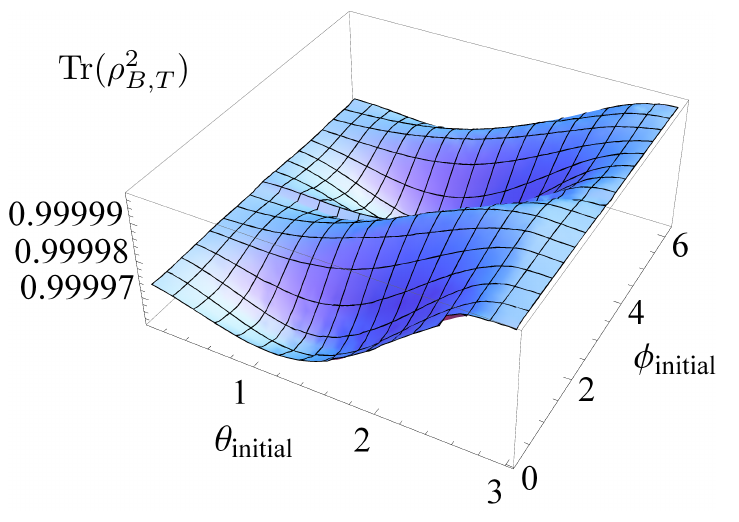}\end{tabular}
  \caption{ For $\lambda|\alpha|=0.01$ ($|\alpha|$ is the amplitude of the $\omega_1$-mode coherent state), and for an initial probe state with $p=\frac{\ii}{\pi}$,  (from left to right) {\bf a)} The change in the target qubit's azimuthal coordinate  $\Delta\phi$  maximized over $a\in[0,2.3]$ and $T\in[0,1.5]$ (natural units) for different initial pure states with all possible $\theta$ and $\phi$ on the Bloch sphere. {\bf b)} Accelerations of the probe atom which maximize $\Delta\phi$. In almost all regions,  relativistic regimes optimize the magnitude of the rotation. {\bf c)} Flying times for the probe atom which maximize $\Delta\phi$. This shows again that when both the probe's acceleration and flying time are relativistic ($aT \sim c$), $\Delta\phi$ attains higher values. {\bf d)} Purity of the target state for the values of the probe's time and acceleration which maximize $\Delta \phi$. The target's state remains mostly pure. Similar results in magnitude are obtained in the maximization of $\Delta \theta$.}
  \label{fig:resultsphi}
\end{figure*}

There is a large amount of freedom in the parameter space to tune up the relativistic trajectory of the probe so that it produces the largest and best controlled rotations on the Bloch sphere.  In Fig (\ref{fig:resultsphi} a) we show the largest rotations on the Bloch sphere that are obtained when we optimize the trajectory of the probe to maximize the rotation in the direction of the azimuthal angle. We obtain equivalent results for the polar angle being the maximum rotation of the same  order of magnitude as $\Delta\phi$, and the parameter optimization yielding very similar results (we include the $\theta$ maximization plot in the appendix). The optimization of the parameters was limited to a narrow  range of times $T\in [0,1.5]$ and accelerations $a\in [0,2.3]$ (in natural units) and only uniformly accelerated motion. We can always find optimal values of $a$ and $T$ such that they yield significant rotations of the target's qubit on the Bloch sphere, either in $\phi$ or $\theta$.  In the density plots in Figs. (\ref{fig:resultsphi} b,c) we display the values of $a$ and $T$ that maximizes the rotations in the target qubit. Remarkably, the optimal values of $a$ and $T$ are commonly on the highest portion of the interval, that is, in the relativistic limit, indicating  that we can perform larger rotations when the probe atom is relativistic (when $aT \sim c$ ).  Furthermore, the flatness of these plots suggests that if we allow for higher values of $a$ and $T$ we could maximize $\Delta\phi$ and $\Delta\theta$ even further: The use of increasingly  relativistic trajectories yields  more efficient motion-based quantum gates (in terms of the number of iterations needed to implement a particular rotation). Note that even though these rotations are individually small, due to the small mixedness introduced per atom, one can compose several of this rotations by sending a continuous beam of atoms through the cavity in a similar way the rotations are composed to produce large gates in \cite{aasen}. 

Note that we are using $aT$ as an estimator of how relativistic the motion of the atom becomes. To connect the estimator used here with  the estimator used in \cite{bruschi,Prdvette}, a straightforward calculation reveals that the relativistic regime $aT \gtrsim c$ is the same relativistic regime as $a L \gtrsim c^2$ if the atom crosses the full cavity and the cavity length is fixed.

We need to assess how much decoherence this process introduces: Since we require a second order calculation for the control terms to appear, we know that the two atoms get entangled \cite{Reznik} and that both atoms get entangled with the quantum field \cite{farming}, thus the final state of the target is non-pure. However, we can see in  Fig. (\ref{fig:resultsphi} d) that the mixedness introduced in the target qubit is 2 orders of magnitude below the magnitude of the rotations, a trend that appears also when the parameters are not optimized to yield maximum rotations.

In order to obtain fully deterministic universal quantum gates, we would like to prepare the setting in such a way that all the remote-controlled rotations on the Bloch sphere were completely independent of the initial state of the target.  In Fig. \ref{fig:resultsphi}, where we just maximize the amplitude of the Bloch sphere rotations, the optimal rotation depends on the initial state of the target.  However, it is possible to optimize the set-up to reduce the dependence of the rotations on the initial state of the target by fine-tuning the parameters of the setting. By varying the probe atom's initial state, acceleration and flying times (but keeping them relativistic) it is possible to find large regions of insensitiveness to the target's initial state which  still enable us to perform independent Bloch sphere rotations. While this is not entirely perfect, note that this is a rather simplified setting: a single probe atom moving through the cavity at relativistic speeds. Considering the effects of arrays or beams of relativistic probe atoms will indeed widen our parameter space and thus our ability to remote-control the target qubit. This will be subject of further study.

{\it Experimental feasibility.-} Although this work mainly focuses on the fundamental proof that the state of a non-relativistic qubit can be fully controlled just by controlling the relativistic motion of a probe atom,  it is relevant to explore and assess the feasibility of possible experimental settings where, in principle, these setups may be implemented. The first aspect we need to consider is the `translation' from the dimensionless unit system (employed in the theoretical discussion) to dimensionful units. The natural scale of units is set by choosing units for $\Omega$, namely $\tilde a=a(\Omega c / \pi)$.   If $\Omega$ lies within the microwave regime (around the order of GHz), one of our dimensionless units of acceleration would correspond to $10^{16}g$ ($g\approx9.8 \text{m}/\text{s}^2$),  so to have non-negligible rotations, as we show on Fig. \ref{fig:results}, we would need accelerations of $\sim10^{14}g$. This is two orders of magnitude below the best previous proposal for detection of the Unruh effect with the same atomic gap \cite{BerryPh,Chen1999a}, one order of magnitude smaller than in the earlier results on relativistic quantum gates \cite{aasen},  and feasible at least in principle \cite{ruso}.

Actually, the required acceleration can be reduced even more by using a detector with a narrower gap. For instance the use of two states connected by a hyperfine transition  or two non-degenerate states of nuclear spin as our qubit  would reduce the energy gap to the order of MHz  \cite{hyperfine1} thus lowering the acceleration threshold to below $\sim10^{11}g$. It is interesting to note that this scale of accelerations for heavy atomic nuclei is already well below the accelerations that can be reached at the LHC \cite{Atlas1}. It would be even possible to think of Stark shifted levels or Zenner-shifts as mechanisms to generate logical qubits with an energy gap in the regime of Hz, thus decreasing the acceleration requirements to below $\sim10^{5}g$. Much closer to experimental  achievability  for the short times required.

There could be some concern regarding the actual implementation of the accelerated probe. Namely,  it is difficult to conceive that such a high acceleration could be sustained for times long enough to allow for a non-negligible rotation, and, more importantly, that the mechanisms that accelerate the probe will not disturb the state of the target or the cavity field. However, as opposed to previous proposals we do not need to accelerate the logical qubit (as in  \cite{aasen}) or the boundary conditions of a cavity (as in \cite{bruschi}). Instead we need to control the trajectory of  a different atomic probe. 

This means that we could conceive accelerating the probe with a laser pulse that has spatial support only in the close proximity of the probe, and that does not cover the spatial region where the target atom is placed in the cavity. Such laser-engineered potentials can, in principle, accelerate neutral atoms to extremely high accelerations (see, for instance \cite{ruso}). Additionally, in order to transmit kinetic energy to the center-of-mass degree of freedom of the probe, one may think of using an additional transition of the probe atom which is largely detuned from the resonant frequency of the first atom, thus minimizing the cross-talk between the interaction used to accelerate the probe and the relevant field modes that serve as a means of communication between target and probe.

Nevertheless, it is much more feasible to think of implementations of quantum simulators that can engineer the relativistic Hamiltonians studied here without the need of physically accelerating a qubit.  Current technology of ion traps and superconducting circuits allows for implementations where relativistic effects can be measured \cite{JleonSabin2a,JleonSabin2b,resin,supercond,DCasimir,PastFutPRL}. More useful for our setting, the simulation of relativistically accelerating atoms in ion traps and superconducting circuits was previously studied in \cite{Diegger}. The family of simulators proposed in \cite{Diegger} are good analogues of the physical setup we study in this paper. As discussed in \cite{Diegger}, current technology in acousto-optical resonators can produce variations of the optical phase in a rate much beyond the required scales to build an experimental realization of what is proposed here. This would be achieved  by means of standard experimental techniques from trapped ion quantum computation  \cite{Leibfried2003}.

Additionally, and as also discussed in \cite{Diegger}, superconducting qubits ultra-strongly coupled to a microwave cavity \cite{Wallraff,Solano} provide a natural setup of an analogue setting where this experiment can be realized. In this case, the relativistic atom Hamiltonian is simulated by means of  the driving of the qubit frequency using the techniques published in \cite{sidebands}.

 Although the scope of this work is to consider the possibility of implementation of single-qubit gates via remote control, it is worth mentioning that it should be possible, in principle, to  implement also  two-qubit gates in this scenario via relativistic effects. Indeed, given that if we can have a controlled generation of entanglement plus the ability to perform universal single qubit gates yields universal 2-qubit quantum gates. In principle, one could combine the entanglement farming techniques developed in \cite{farming} (also based in the motion of atomic probes in optical cavities) with the relativistic universal one-qubit gates proposed here to yield an universal 2-qubit gates. Another possibility well as with methods based on moving boundary conditions to implement Gaussian gates \cite{bruschi}. This would yield universal quantum gates exclusively based on motion and relativistic effects. Of course there is a number of possible avenues to implement the entanglement generation needed to propose two-qubit gates.  For example, there is evidence that a non-trivial entangling unitary on two qubits is obtained already by putting them simultaneously in the cavity \cite{farming,supercond,DCasimir,PastFutPRL}. Or also, if the qubits enter the cavity sequentially, the first qubit going through the cavity will modify the field state such that the second qubit may be excited or not, depending on the state of the first qubit. The transformation of the second qubit will depend on the state of the first qubit (similar to a CNOT gate) and that would be dependent on the state of motion of that qubit. These are out of the scope of the current paper but will be considered elsewhere.

{\it Conclusions.-} We showed that it is possible to remote control a stationary atomic qubit by controlling the relativistic motion of a different probe atom. We verified that we can obtain independent Bloch sphere rotations on the target qubit with negligible mixedness by controlling the non-inertial trajectory  of the probe.

Remarkably,  the more relativistic the motion of the probe the more efficient the remote manipulation of the target's internal state. This suggests that relativistic effects can be thought of as a resource for quantum control.  Although clearly  this scheme may, at its current stage, is obviously not suited (in term of resources) to compete with the usual approaches to the implementation ofr quantum gates, it does indeed highlight a connection of quantum information with high-energy physics, possibly connecting the relativistic trajectory of accelerated nuclei at energy scales within reach of current particle colliders with operations on stationary qubits. 

Furthermore, this fundamental study reveals that universal quantum gates based uniquely on relativistic motion of an atomic probe may be implemented and/or simulated (e.g. in superconducting circuit settings). The concrete study of the parameter variation and error tolerance of such a setting to produce universal gates is currently a subject of study and will soon appear elsewhere. We believe that this constitutes another step towards understanding in what scenarios relativistic approaches to quantum information technologies may be beneficial.

{\it Acknowledgements.- } E. M-M. acknowledges support of the Banting Postdoctoral Fellowship Programme.

\appendix

\section{APPENDIX: Detail of the second order time evolution calculation}
The main text details that the interaction Hamiltonian (in the interaction picture) of the probe atom, the target qubit and the quantum field in the Dirichlet cavity is given by
 \begin{equation}
 H_I(t)=\frac{\d \tau}{\d t} H^{(\text{A})}_{I}[\tau(t)]+H^{(\text{B})}_{I}(t),
 \end{equation}
where the individual $H^{(\text{D})}_{I}$ are the Unruh-Dewitt Hamiltonians of the individual detectors.

Since we are working in the stationary rest frame of the target, the redshift factor $\frac{d\tau}{dt}$ coming from the different proper times of the probe and the target will be absorbed into the probe's switching function  $\xi_{A} (t)$ for convenience. Note that in this case $\tau_{B}(t)=t$, whereas $t(\tau_A)=a^{-1}\sinh a\tau_A$, $x(\tau_A)=a^{-1}(\cosh a\tau_A -1)$. Considering all this, $U^{(1)}$ can be written more explicitly as 
\begin{align}
& \nonumber U^{(1)}= -\ii\lambda_{A}\sum_{j}(I_{+,j}^{(A)}a_{j}^{\dagger}\sigma^{+}_{A}+I_{-,j}^{(A)}a_{j}^{\dagger}\sigma^{-}_{A}+I_{-,j}^{(A)*}a_{j}\sigma^{+}_{A}\\
& \nonumber +I_{+,j}^{(A)*}a_{j}\sigma^{-}_{A}) -\ii\lambda_{B}\sum_{j}(I_{+,j}^{(B)}a_{j}^{\dagger}\sigma^{+}_{B}+I_{-,j}^{(B)}a_{j}^{\dagger}\sigma^{-}_{B}+\\
& I_{-,j}^{(B)*}a_{j}\sigma^{+}_{B}+I_{+,j}^{(B)*}a_{j}\sigma^{-}_{B})
\end{align}
Where the integrals $I_{\pm,j}^{\nu}$ are given in the main text. In order to obtain the perturbative corrections to the target qubit's final state, we must trace out the probe system and the quantum field, which yields  
\begin{align}
 \nonumber \rho^{(1)}_{T,B}&=\text{Tr}_{A}(\text{Tr}_{f}(U^{(1)}\rho_{0}))+\text{Tr}_{A}(\text{Tr}_{f}(\rho_{0}U^{(1)\dagger}))\\
 \nonumber &=-\ii\lambda_{B}(\alpha^{*}(I_{+,1}^{(B)}\sigma^{+}_{B}+I_{-,1}^{(B)}\sigma^{-}_{B}) \\
&\;\;\;\;+ \alpha(I_{-,1}^{(B)*}\sigma^{+}_{B}+I_{+,1}^{(B)*}\sigma^{-}_{B}))
\rho_{B,0}-\text{H.c.}.
\label{eq:firstorder}
\end{align}
Recall it is the second order terms ($\lambda_{A}\lambda_{B}$) that allow remote control to be obtained on the target qubit in the cavity.

 It is convenient for the second order calculation to explicitly write the target and probe qubit's initial states in the general form
\begin{equation}
\rho_{0}^{\text{B}}=
\begin{pmatrix}
\varphi & \delta\\
\delta^{*} & \kappa
\end{pmatrix},\qquad
\rho_{0}^{\text{A}}=
\begin{pmatrix}
\eta & \gamma\\
\gamma^{*} & \beta
\end{pmatrix},
\end{equation}
where, of course $\beta=1-\eta$ is real and the $\rho_0^{A,B}$ are positive. Notice that in our calculations we took the state of the probe to be pure: $\ket{\psi_A}=p\ket{g}+\sqrt{1-p^2}\ket{e}$. This would mean that $\eta=p^2,\beta=1-p^2,\gamma=p\sqrt{1-p^2}$.  The fundamental mode of the cavity is prepared in a coherent state $\ket{\alpha}$. This makes the expressions for the second order corrections longer because $\alpha$ multiplies only the first term in the sum over the field modes, and so, each sum is split into two terms. To simplify notation, we will make use of the labels $-A$ or $-B$ meaning $\Omega_{A}\rightarrow -\Omega_{A}$ and $\Omega_{B}\rightarrow -\Omega_{B}$ respectively. Now we detail the calculation of the target qubit's final state up to second order in perturbation theory. To make it easier for the reader to repeat all the calculations,  let us write all the terms in the second order time evolution contribution explicitly.  First we apply $U^{(2)}$ to $\rho_{0}$ to obtain
\begin{align}
\nonumber U^{(2)}\rho_{0}=-\Big[&\nonumber \sum_{j}\lambda_{A}^2[J_{+,+,j}^{A,A}(\sigma_{A}^{+})^{2}(a_{j}^{\dagger})^{2}+J_{+,-,j}^{A,A}(\sigma_{A}^{+})^{2}a_{j}^{\dagger}a_{j}\\[-1.5mm]
&\nonumber + J_{-,+,j}^{A,A}\sigma_{A}^{+}\sigma_{A}^{-}(a_{j}^{\dagger})^{2}+J_{-,-,j}^{A,A}\sigma_{A}^{+}\sigma_{A}^{-}a_{j}^{\dagger}a_{j}\\
&\nonumber +J_{-,-,j}^{-A,A*}(\sigma_{A}^{\dagger})^{2}a_{j}a_{j}^{\dagger}+J_{-,+,j}^{-A,A*}(\sigma_{A}^{\dagger})^{2}a_{j}^{2}\\
&\nonumber  +J_{+,-,j}^{-A,A*}\sigma_{A}^{+}\sigma_{A}^{-}a_{j}a_{j}^{\dagger}+J_{+,+,j}^{-A,A*}\sigma_{A}^{+}\sigma_{A}^{-}a_{j}a_{j}\\
&\nonumber +J_{+,+,j}^{-A,A}\sigma_{A}^{-}\sigma_{A}^{+}(a_{j}^{\dagger})^{2}+J_{+,-,j}^{-A,A}\sigma_{A}^{-}\sigma_{A}^{+}a_{j}^{\dagger}a_{j}\\
&\nonumber +J_{-,+,j}^{-A,A}(\sigma_{A}^{-})^{2}(a_{j}^{\dagger})^{2}+J_{-,-,j}^{-A,A}(\sigma_{A}^{-})^{2}a_{j}^{\dagger}a_{j}+\\
&\nonumber +J_{-,-,j}^{A,A*}\sigma_{A}^{-}\sigma_{A}^{+}a_{j}a_{j}^{\dagger}+J_{-,+,j}^{A,A*}\sigma_{A}^{-}\sigma_{A}^{+}a_{j}^{2}\\
&\nonumber +J_{+,-,j}^{A,A*}(\sigma_{A}^{-})^{2}a_{j}a_{j}^{\dagger}+J_{+,+,j}^{A,A*}(\sigma_{A}^{-})^{2}a_{j}^{2}]\\
&\nonumber +\lambda_{A}\lambda_{B}[J_{+,+,j}^{A,B}\sigma_{A}^{+}\sigma_{B}^{+}(a_{j}^{\dagger})^{2}+J_{+,-,j}^{A,B}\sigma_{A}^{+}\sigma_{B}^{+}a_{j}^{\dagger}a_{j}\\
&\nonumber +J_{-,+,j}^{A,B}\sigma_{A}^{+}\sigma_{B}^{-}(a_{j}^{\dagger})^{2}+J_{-,-,j}^{A,B}\sigma_{A}^{+}\sigma_{B}^{-}a_{j}^{\dagger}a_{j}\\
&\nonumber +J_{-,-,j}^{-A,B*}\sigma_{A}^{+}\sigma_{B}^{+}a_{j}a_{j}^{\dagger}+J_{-,+,j}^{-A,B*}\sigma_{A}^{+}\sigma_{B}^{+}(a_{j})^{2}\\
&\nonumber +J_{+,-,j}^{-A,B*}\sigma_{A}^{+}\sigma_{B}^{-}a_{j}a_{j}^{\dagger}+J_{+,+,j}^{-A,B*}\sigma_{A}^{+}\sigma_{B}^{-}(a_{j}^{2})\\
&\nonumber +J_{+,+,j}^{-A,B}\sigma_{A}^{-}\sigma_{B}^{+}(a_{j}^{\dagger})^{2}+J_{+,-,j}^{-A,B}\sigma_{A}^{-}\sigma_{B}^{+}a_{j}^{\dagger}a_{j}\\
&\nonumber +J_{-,+,j}^{-A,B}\sigma_{A}^{-}\sigma_{B}^{-}(a_{j}^{\dagger})^{2}+J_{-,-,j}^{-A,B}\sigma_{A}^{-}\sigma_{B}^{-}a_{j}^{\dagger}a_{j}\\
&\nonumber +J_{-,-,j}^{A,B*}\sigma_{A}^{-}\sigma_{B}^{+}a_{j}a_{j}^{\dagger}+J_{-,+,j}^{A,B*}\sigma_{A}^{-}\sigma_{B}^{+}a_{j}^{2}\\
&\nonumber +J_{+,-,j}^{A,B*}\sigma_{A}^{-}\sigma_{B}^{-}a_{j}a_{j}^{\dagger}+J_{+,+,j}^{A,B*}\sigma_{A}^{-}\sigma_{B}^{-}(a_{j})^{2}]\\
&\nonumber +\lambda_{B}\lambda_{A}[J_{+,+,j}^{B,A}\sigma_{B}^{+}\sigma_{A}^{+}(a_{j}^{\dagger})^{2}+J_{+,-,j}^{B,A}\sigma_{B}^{+}\sigma_{A}^{+}a_{j}^{\dagger}a_{j}\\
&\nonumber +J_{-,+,j}^{B,A}\sigma_{B}^{+}\sigma_{A}^{-}(a_{j}^{\dagger})^{2}+J_{-,-,j}^{B,A}\sigma_{B}^{+}\sigma_{A}^{-}a_{j}^{\dagger}a_{j}\\
&\nonumber +J_{-,-,j}^{-B,A*}\sigma_{B}^{+}\sigma_{A}^{+}a_{j}a_{j}^{\dagger}+J_{-,+,j}^{-B,A*}\sigma_{B}^{+}\sigma_{A}^{+}a_{j}^{2}\\
&\nonumber +J_{+,-,j}^{-B,A*}\sigma_{B}^{+}\sigma_{A}^{-}a_{j}a_{j}^{\dagger}+J_{+,+,j}^{-B,A*}\sigma_{B}^{+}\sigma_{A}^{-}a_{j}^{2}\\
&\nonumber +J_{+,+,j}^{-B,A}\sigma_{B}^{-}\sigma_{A}^{+}(a_{j}^{\dagger})^{2}+J_{+,-,j}^{-B,A}\sigma_{B}^{-}\sigma_{A}^{+}a_{j}^{\dagger}a_{j}\\
&\nonumber +J_{-,+,j}^{-B,A}\sigma_{B}^{-}\sigma_{A}^{-}(a_{j}^{\dagger})^{2}+J_{-,-,j}^{-B,A}\sigma_{B}^{-}\sigma_{A}^{-}a_{j}^{\dagger}a_{j}\\
&\nonumber +J_{-,-,j}^{B,A*}\sigma_{B}^{-}\sigma_{A}^{+}a_{j}a_{j}^{\dagger}+J_{-,+,j}^{B,A*}\sigma_{B}^{-}\sigma_{A}^{+}a_{j}^{2}\\
&\nonumber +J_{+,-,j}^{B,A*}\sigma_{B}^{-}\sigma_{A}^{-}a_{j}a_{j}^{\dagger}+J_{+,+,j}^{B,A*}\sigma_{B}^{-}\sigma_{A}^{-}a_{j}^{2}]\\
&\nonumber +\lambda_{B}^{2}[J_{+,+,j}^{B,B}(\sigma_{B}^{+})^{2}(a_{j}^{\dagger})^{2}+J_{+,-,j}^{B,B}(\sigma_{B}^{+})^{2}a_{j}^{\dagger}a_{j}\\
&\nonumber +J_{-,+,j}^{B,B}\sigma_{B}^{+}\sigma_{B}^{-}(a_{j}^{\dagger})^{2}+J_{-,-,j}^{B,B}\sigma_{B}^{+}\sigma_{B}^{-}a_{j}^{\dagger}a_{j}\\
&\nonumber +J_{-,-,j}^{-B,B*}(\sigma_{B}^{+})^{2}a_{j}a_{j}^{\dagger}+J_{-,+,j}^{-B,B*}(\sigma_{B}^{+})^{2}a_{j}^{2}\\
&\nonumber +J_{+,-,j}^{-B,B*}\sigma_{B}^{+}\sigma_{B}^{-}a_{j}a_{j}^{\dagger}+J_{+,+,j}^{-B,B*}\sigma_{B}^{+}\sigma_{B}^{-}a_{j}^{2}\\
&\nonumber +J_{+,+,j}^{-B,B}\sigma_{B}^{-}\sigma_{B}^{+}(a_{j}^{\dagger})^{2}+J_{+,-,j}^{-B,B}\sigma_{B}^{-}\sigma_{B}^{+}a_{j}^{\dagger}a_{j}\\
&\nonumber +J_{-,+,j}^{-B,B}(\sigma_{B}^{-})^{2}(a_{j}^{\dagger})^{2}+J_{-,-,j}^{-B,B}(\sigma_{B}^{-})^{2}a_{j}^{\dagger}a_{j}\\
&\nonumber +J_{-,-,j}^{B,B*}\sigma_{B}^{-}\sigma_{B}^{+}a_{j}a_{j}^{\dagger}+J_{-,+,j}^{B,B*}\sigma_{B}^{-}\sigma_{B}^{+}a_{j}^{2}\\
& +J_{+,-,j}^{B,B*}(\sigma_{B}^{-})^{2}a_{j}a_{j}^{\dagger}+J_{+,+,j}^{B,B*}(\sigma_{B}^{-})^{2}a_{j}^{2}]\Big]\rho_{0}.
\end{align}
Where  the $J_{\pm,\pm,j}^{\nu,\mu}$ integrals are given in the main text. Recall that we are interested in the target qubit's  final state $\text{Tr}_{A}[\text{Tr}_{f}(\rho_{T}^{2})]$; thus the next step is to trace out the field from the above equation to obtain
\begin{align}
\nonumber \text{Tr}_{f}&(U^{(2)}\rho_{0})=
-[\lambda_{A}^2[J_{+,+,1}^{A,A}(\sigma_{A}^{+})^{2}(\alpha^{*})^{2}+J_{+,-,1}^{A,A}(\sigma_{A}^{+})^{2}|\alpha|^{2}\\
&\nonumber +J_{-,+,1}^{A,A}\sigma_{A}^{+}\sigma_{A}^{-}(\alpha^{*})^{2}+J_{-,-,1}^{A,A}\sigma_{A}^{+}\sigma_{A}^{-}|\alpha|^{2}\\
&\nonumber +J_{-,-,1}^{-A,A*}(\sigma_{A}^{\dagger})^{2}(1+|\alpha|^{2})+J_{-,+,1}^{-A,A*}(\sigma_{A}^{\dagger})^{2}\alpha^{2}\\
&\nonumber +J_{+,-,1}^{-A,A*}\sigma_{A}^{+}\sigma_{A}^{-}(1+|\alpha|^{2})+J_{+,+,1}^{-A,A*}\sigma_{A}^{+}\sigma_{A}^{-}(\alpha)^{2}\\
&\nonumber +J_{+,+,1}^{-A,A}\sigma_{A}^{-}\sigma_{A}^{+}(\alpha^{*})^{2}+J_{+,-,1}^{-A,A}\sigma_{A}^{-}\sigma_{A}^{+}|\alpha|^{2}\\
&\nonumber +J_{-,+,1}^{-A,A}(\sigma_{A}^{-})^{2}(\alpha^{*})^{2}+J_{-,-,1}^{-A,A}(\sigma_{A}^{-})^{2}|\alpha|^{2}\\
&\nonumber +J_{-,-,1}^{A,A*}\sigma_{A}^{-}\sigma_{A}^{+}(1+|\alpha|^{2})+J_{-,+,1}^{A,A*}\sigma_{A}^{-}\sigma_{A}^{+}\alpha^{2}\\
&\nonumber +J_{+,-,1}^{A,A*}(\sigma_{A}^{-})^{2}(1+|\alpha|^{2})+J_{+,+,1}^{A,A*}(\sigma_{A}^{-})^{2}\alpha^{2}]\\
&\nonumber +\lambda_{A}\lambda_{B}[J_{+,+,1}^{A,B}\sigma_{A}^{+}\sigma_{B}^{+}(\alpha^{*})^{2}+J_{+,-,1}^{A,B}\sigma_{A}^{+}\sigma_{B}^{+}|\alpha|^{2}\\
&\nonumber +J_{-,+,1}^{A,B}\sigma_{A}^{+}\sigma_{B}^{-}(\alpha^{*})^{2}+J_{-,-,1}^{A,B}\sigma_{A}^{+}\sigma_{B}^{-}|\alpha|^{2}\\
&\nonumber +J_{-,-,1}^{-A,B*}\sigma_{A}^{+}\sigma_{B}^{+}(1+|\alpha|^{2})+J_{-,+,1}^{-A,B*}\sigma_{A}^{+}\sigma_{B}^{+}\alpha^{2}\\
&\nonumber +J_{+,-,1}^{-A,B*}\sigma_{A}^{+}\sigma_{B}^{-}(1+|\alpha|^{2})+J_{+,+,1}^{-A,B*}\sigma_{A}^{+}\sigma_{B}^{-}\alpha^{2}\\
&\nonumber +J_{+,+,1}^{-A,B}\sigma_{A}^{-}\sigma_{B}^{+}(\alpha^{*})^{2}+J_{+,-,1}^{-A,B}\sigma_{A}^{-}\sigma_{B}^{+}|\alpha|^{2}\\
&\nonumber +J_{-,+,1}^{-A,B}\sigma_{A}^{-}\sigma_{B}^{-}(\alpha^{*})^{2}+J_{-,-,1}^{-A,B}\sigma_{A}^{-}\sigma_{B}^{-}|\alpha|^{2}\\
&\nonumber +J_{-,-,1}^{A,B*}\sigma_{A}^{-}\sigma_{B}^{+}(1+|\alpha|^{2})+J_{-,+,1}^{A,B*}\sigma_{A}^{-}\sigma_{B}^{+}\alpha^{2}\\
&\nonumber +J_{+,-,1}^{A,B*}\sigma_{A}^{-}\sigma_{B}^{-}(1+|\alpha|^{2})+J_{+,+,j}^{A,B*}\sigma_{A}^{-}\sigma_{B}^{-}\alpha^{2}]\\
&\nonumber +\lambda_{B}\lambda_{A}[J_{+,+,1}^{B,A}\sigma_{B}^{+}\sigma_{A}^{+}(\alpha^{*})^{2}+J_{+,-,1}^{B,A}\sigma_{B}^{+}\sigma_{A}^{+}|\alpha|^{2}\\
&\nonumber +J_{-,+,1}^{B,A}\sigma_{B}^{+}\sigma_{A}^{-}(\alpha^{*})^{2}+J_{-,-,1}^{B,A}\sigma_{B}^{+}\sigma_{A}^{-}|\alpha|^{2}\\
&\nonumber +J_{-,-,1}^{-B,A*}\sigma_{B}^{+}\sigma_{A}^{+}(1+|\alpha|^{2})+J_{-,+,1}^{-B,A*}\sigma_{B}^{+}\sigma_{A}^{+}\alpha^{2}\\
&\nonumber +J_{+,-,1}^{-B,A*}\sigma_{B}^{+}\sigma_{A}^{-}(1+|\alpha|^{2})+J_{+,+,1}^{-B,A*}\sigma_{B}^{+}\sigma_{A}^{-}\alpha^{2}\\
&\nonumber +J_{+,+,1}^{-B,A}\sigma_{B}^{-}\sigma_{A}^{+}(\alpha^{*})^{2}+J_{+,-,1}^{-B,A}\sigma_{B}^{-}\sigma_{A}^{+}|\alpha|^{2}\\
&\nonumber +J_{-,+,1}^{-B,A}\sigma_{B}^{-}\sigma_{A}^{-}(\alpha^{*})^{2}+J_{-,-,1}^{-B,A}\sigma_{B}^{-}\sigma_{A}^{-}|\alpha|^{2}\\
&\nonumber +J_{-,-,1}^{B,A*}\sigma_{B}^{-}\sigma_{A}^{+}(1+|\alpha|^{2})+J_{-,+,1}^{B,A*}\sigma_{B}^{-}\sigma_{A}^{+}\alpha^{2}\\
&\nonumber +J_{+,-,1}^{B,A*}\sigma_{B}^{-}\sigma_{A}^{-}(1+|\alpha|^{2})+J_{+,+,1}^{B,A*}\sigma_{B}^{-}\sigma_{A}^{-}\alpha^{2}]\\
&\nonumber +\lambda_{B}^{2}[J_{+,+,1}^{B,B}(\sigma_{B}^{+})^{2}(\alpha^{*})^{2}+J_{+,-,1}^{B,B}(\sigma_{B}^{+})^{2}|\alpha|^{2}\\
&\nonumber +J_{-,+,1}^{B,B}\sigma_{B}^{+}\sigma_{B}^{-}(\alpha^{*})^{2}+J_{-,-,1}^{B,B}\sigma_{B}^{+}\sigma_{B}^{-}|\alpha|^{2}\\
&\nonumber +J_{-,-,1}^{-B,B*}(\sigma_{B}^{+})^{2}(1+|\alpha|^{2})+J_{-,+,1}^{-B,B*}(\sigma_{B}^{+})^{2}\alpha^{2}\\
&\nonumber +J_{+,-,1}^{-B,B*}\sigma_{B}^{+}\sigma_{B}^{-}(1+|\alpha|^{2})+J_{+,+,j}^{-B,B*}\sigma_{B}^{+}\sigma_{B}^{-}\alpha^{2}\\
&\nonumber +J_{+,+,1}^{-B,B}\sigma_{B}^{-}\sigma_{B}^{+}(\alpha^{*})^{2}+J_{+,-,1}^{-B,B}\sigma_{B}^{-}\sigma_{B}^{+}|\alpha|^{2}\\
&\nonumber +J_{-,+,1}^{-B,B}(\sigma_{B}^{-})^{2}(\alpha^{*})^{2}+J_{-,-,1}^{-B,B}(\sigma_{B}^{-})^{2}|\alpha|^{2}\\
&\nonumber +J_{-,-,1}^{B,B*}\sigma_{B}^{-}\sigma_{B}^{+}(1+|\alpha|^{2})+J_{-,+,1}^{B,B*}\sigma_{B}^{-}\sigma_{B}^{+}\alpha^{2}\\
&\nonumber +J_{+,-,1}^{B,B*}(\sigma_{B}^{-})^{2}(1+|\alpha|^{2})+J_{+,+,1}^{B,B*}(\sigma_{B}^{-})^{2}\alpha^{2}]]\rho_{0}^{\text{A}}\otimes\rho_{0}^{\text{B}}\\
&\nonumber -\sum_{j=2}^{n}[\lambda_{A}^{2}[J_{-,-,j}^{-A,A*}(\sigma_{A}^{\dagger})^{2}+J_{+,-,j}^{-A,A*}\sigma_{A}^{+}\sigma_{A}^{-}\\
&\nonumber +J_{-,-,j}^{A,A*}\sigma_{A}^{-}\sigma_{A}^{+}+J_{+,-,j}^{A,A*}(\sigma_{A}^{-})^{2}]\\
&\nonumber +\lambda_{A}\lambda_{B}[J_{-,-,j}^{-A,B*}\sigma_{A}^{+}\sigma_{B}^{+}+J_{+,-,j}^{-A,B*}\sigma_{A}^{+}\sigma_{B}^{-}\\
&\nonumber +J_{-,-,j}^{A,B*}\sigma_{A}^{-}\sigma_{B}^{+}+J_{+,-,j}^{A,B*}\sigma_{A}^{-}\sigma_{B}^{-}]\\
&\nonumber +\lambda_{A}\lambda_{B}[J_{-,-,j}^{-B,A*}\sigma_{B}^{+}\sigma_{A}^{+}+J_{+,-,j}^{-B,A*}\sigma_{B}^{+}\sigma_{A}^{-}\\
&\nonumber +J_{-,-,j}^{B,A*}\sigma_{B}^{-}\sigma_{A}^{+}+J_{+,-,j}^{B,A*}\sigma_{B}^{-}\sigma_{A}^{-}]\\
&\nonumber +\lambda_{B}^{2}[J_{-,-,j}^{-B,B*}(\sigma_{B}^{+})^{2}+J_{+,-,j}^{-B,B*}\sigma_{B}^{+}\sigma_{B}^{-}\\
& +J_{-,-,j}^{B,B*}\sigma_{B}^{-}\sigma_{B}^{+}+J_{+,-,j}^{B,B*}(\sigma_{B}^{-})^{2}]]\rho_{0}^{\text{A}}\otimes\rho_{0}^{\text{B}},
\end{align}
where we have used the fact that $(\sigma_{d}^{\pm})^{2}=0$. Now we trace out over the probe atom's degrees of freedom to obtain
\begin{align}
\nonumber \text{Tr}_{A}&(\text{Tr}_{f}(U^{(2)}\rho_{0}))=
-[\lambda_{A}^2[J_{-,+,1}^{A,A}\eta(\alpha^{*})^{2}+J_{-,-,1}^{A,A}\eta|\alpha|^{2}\\
&\nonumber +J_{+,-,1}^{-A,A*}\eta(1+|\alpha|^{2})+J_{+,+,1}^{-A,A*}\eta(\alpha)^{2}+\\
&\nonumber J_{+,+,1}^{-A,A}\beta(\alpha^{*})^{2}+J_{+,-,1}^{-A,A}\beta|\alpha|^{2}\\
&\nonumber +J_{-,-,1}^{A,A*}\beta(1+|\alpha|^{2})+J_{-,+,1}^{A,A*}\beta\alpha^{2}]\\
&\nonumber +\lambda_{A}\lambda_{B}[J_{+,+,1}^{A,B}\gamma^{*}\sigma_{B}^{+}(\alpha^{*})^{2}+J_{+,-,1}^{A,B}\gamma^{*}\sigma_{B}^{+}|\alpha|^{2}\\
&\nonumber +J_{-,+,1}^{A,B}\gamma^{*}\sigma_{B}^{-}(\alpha^{*})^{2}+J_{-,-,1}^{A,B}\gamma^{*}\sigma_{B}^{-}|\alpha|^{2}\\
&\nonumber +J_{-,-,1}^{-A,B*}\gamma^{*}\sigma_{B}^{+}(1+|\alpha|^{2})+J_{-,+,1}^{-A,B*}\gamma^{*}\sigma_{B}^{+}\alpha^{2}\\
&\nonumber +J_{+,-,1}^{-A,B*}\gamma^{*}\sigma_{B}^{-}(1+|\alpha|^{2})+J_{+,+,1}^{-A,B*}\gamma^{*}\sigma_{B}^{-}\alpha^{2}\\
&\nonumber +J_{+,+,1}^{-A,B}\gamma\sigma_{B}^{+}(\alpha^{*})^{2}+J_{+,-,1}^{-A,B}\gamma\sigma_{B}^{+}|\alpha|^{2}\\
&\nonumber +J_{-,+,1}^{-A,B}\gamma\sigma_{B}^{-}(\alpha^{*})^{2}+J_{-,-,1}^{-A,B}\gamma\sigma_{B}^{-}|\alpha|^{2}\\
&\nonumber +J_{-,-,1}^{A,B*}\gamma\sigma_{B}^{+}(1+|\alpha|^{2})+J_{-,+,1}^{A,B*}\gamma\sigma_{B}^{+}\alpha^{2}\\
&\nonumber +J_{+,-,1}^{A,B*}\gamma\sigma_{B}^{-}(1+|\alpha|^{2})+J_{+,+,j}^{A,B*}\gamma\sigma_{B}^{-}\alpha^{2}]\\
&\nonumber +\lambda_{B}\lambda_{A}[J_{+,+,1}^{B,A}\sigma_{B}^{+}\gamma^{*}(\alpha^{*})^{2}+J_{+,-,1}^{B,A}\sigma_{B}^{+}\gamma^{*}|\alpha|^{2}\\
&\nonumber +J_{-,+,1}^{B,A}\sigma_{B}^{+}\gamma(\alpha^{*})^{2}+J_{-,-,1}^{B,A}\sigma_{B}^{+}\gamma|\alpha|^{2}\\
&\nonumber +J_{-,-,1}^{-B,A*}\sigma_{B}^{+}\gamma^{*}(1+|\alpha|^{2})+J_{-,+,1}^{-B,A*}\sigma_{B}^{+}\gamma^{*}\alpha^{2}\\
&\nonumber +J_{+,-,1}^{-B,A*}\sigma_{B}^{+}\gamma(1+|\alpha|^{2})+J_{+,+,1}^{-B,A*}\sigma_{B}^{+}\gamma\alpha^{2}\\
&\nonumber +J_{+,+,1}^{-B,A}\sigma_{B}^{-}\gamma^{*}(\alpha^{*})^{2}+J_{+,-,1}^{-B,A}\sigma_{B}^{-}\gamma^{*}|\alpha|^{2}\\
&\nonumber +J_{-,+,1}^{-B,A}\sigma_{B}^{-}\gamma(\alpha^{*})^{2}+J_{-,-,1}^{-B,A}\sigma_{B}^{-}\gamma|\alpha|^{2}\\
&\nonumber +J_{-,-,1}^{B,A*}\sigma_{B}^{-}\gamma^{*}(1+|\alpha|^{2})+J_{-,+,1}^{B,A*}\sigma_{B}^{-}\gamma^{*}\alpha^{2}\\
&\nonumber +J_{+,-,1}^{B,A*}\sigma_{B}^{-}\gamma(1+|\alpha|^{2})+J_{+,+,1}^{B,A*}\sigma_{B}^{-}\gamma\alpha^{2}]\\
&\nonumber +\lambda_{B}^{2}[J_{-,+,1}^{B,B}\sigma_{B}^{+}\sigma_{B}^{-}(\alpha^{*})^{2}+J_{-,-,1}^{B,B}\sigma_{B}^{+}\sigma_{B}^{-}|\alpha|^{2}\\
&\nonumber +J_{+,-,1}^{-B,B*}\sigma_{B}^{+}\sigma_{B}^{-}(1+|\alpha|^{2})+J_{+,+,j}^{-B,B*}\sigma_{B}^{+}\sigma_{B}^{-}\alpha^{2}\\
&\nonumber +J_{+,+,1}^{-B,B}\sigma_{B}^{-}\sigma_{B}^{+}(\alpha^{*})^{2}+J_{+,-,1}^{-B,B}\sigma_{B}^{-}\sigma_{B}^{+}|\alpha|^{2}\\
&\nonumber +J_{-,-,1}^{B,B*}\sigma_{B}^{-}\sigma_{B}^{+}(1+|\alpha|^{2})+J_{-,+,1}^{B,B*}\sigma_{B}^{-}\sigma_{B}^{+}\alpha^{2}
]]\rho_{0}^{\text{A}}\otimes\rho_{0}^{\text{B}}\\
&\nonumber -\sum_{j=2}^{n}[\lambda_{A}^{2}[J_{+,-,j}^{-A,A*}\eta+J_{-,-,j}^{A,A*}\beta]\\
&\nonumber +\lambda_{A}\lambda_{B}[J_{-,-,j}^{-A,B*}\gamma^{*}\sigma_{B}^{+}+J_{+,-,j}^{-A,B*}\gamma^{*}\sigma_{B}^{-}\\
&\nonumber +J_{-,-,j}^{A,B*}\gamma\sigma_{B}^{+}+J_{+,-,j}^{A,B*}\gamma\sigma_{B}^{-}]\\
&\nonumber +\lambda_{A}\lambda_{B}[J_{-,-,j}^{-B,A*}\sigma_{B}^{+}\gamma^{*}+J_{+,-,j}^{-B,A*}\sigma_{B}^{+}\gamma\\
&\nonumber +J_{-,-,j}^{B,A*}\sigma_{B}^{-}\gamma^{*}+J_{+,-,j}^{B,A*}\sigma_{B}^{-}\gamma]\\
& +\lambda_{B}^{2}[J_{+,-,j}^{-B,B*}\sigma_{B}^{+}\sigma_{B}^{-}+J_{-,-,j}^{B,B*}\sigma_{B}^{-}\sigma_{B}^{+}]]\rho_{0}^{\text{A}}\otimes\rho_{0}^{\text{B}}.
\label{eq:secondorder1}
\end{align}
Now we must also compute $U^{(1)}\rho_{0}U^{(1)\dagger}$:
\begin{align}
&\nonumber U^{(1)}\rho_{0}U^{(1)\dagger}=\\
&\nonumber -\sum_{i,j}[\lambda_{A}(I_{+,j}^{(A)}a_{j}^{\dagger}\sigma^{+}_{A}\!+\!I_{-,j}^{(A)}a_{j}^{\dagger}\sigma^{-}_{A}\!+\!I_{-,j}^{(A)*}a_{j}\sigma^{+}_{A}\!+\!I_{+,j}^{(A)*}a_{j}\sigma^{-}_{A})\\
&\nonumber +\lambda_{B}(I_{+,j}^{(B)}a_{j}^{\dagger}\sigma^{+}_{B}\!+\!I_{-,j}^{(B)}a_{j}^{\dagger}\sigma^{-}_{B}\!+\!I_{-,j}^{(B)*}a_{j}\sigma^{+}_{B}\!+\!I_{+,j}^{(B)*}a_{j}\sigma^{-}_{B})]\rho_{0}\\
&\nonumber [\lambda_{A}(I_{+,i}^{(A)}a_{i}^{\dagger}\sigma^{+}_{A}+I_{-,i}^{(A)}a_{i}^{\dagger}\sigma^{-}_{A}+I_{-,i}^{(A)*}a_{i}\sigma^{+}_{A}+I_{+,i}^{(A)*}a_{i}\sigma^{-}_{A})\\
& +\lambda_{B}(I_{+,i}^{(B)}a_{i}^{\dagger}\sigma^{+}_{B}+I_{-,i}^{(B)}a_{i}^{\dagger}\sigma^{-}_{B}+I_{-,i}^{(B)*}a_{i}\sigma^{+}_{B}+I_{+,i}^{(B)*}a_{i}\sigma^{-}_{B})].
\end{align}
Following the usual procedure, we now trace out the field modes and the probe atom's quantum state to obtain
\begin{align}
&\nonumber \text{Tr}_{A}(\text{Tr}_{f}(U^{(1)}\rho_{0}U^{(1)\dagger}))=\\
&\nonumber \sum_{j=2}^{n}-[\lambda_{A}^{2}[I_{+,1}^{(A)}I_{-,1}^{(A)}(\alpha^{*})^{2}\beta+|I_{+,j}^{(A)}|^{2}\beta+|I_{+,1}^{(A)}|^{2}(1+|\alpha|^{2})\beta\\
&\nonumber + I_{-,1}^{(A)}I_{+,1}^{(A)}(\alpha^{*})^{2}\eta+|I_{-,1}^{(A)}|^{2}(1+|\alpha|^{2})\eta+ |I_{-,j}^{(A)}|^{2}\eta\\
&\nonumber +|I_{-,1}^{(A)}|^{2}|\alpha|^{2}\beta+I_{-,1}^{(A)*}I_{+,1}^{(A)*}\alpha^{2}\beta+|I_{+,1}^{(A)}|^{2}|\alpha|^{2}\eta\\
&\nonumber +I_{+,1}^{(A)*}I_{-,1}^{(A)*}\alpha^{2}\eta]]\rho_{0,B}\\
&\nonumber -\rho_{0,B}[\lambda_{A}\lambda_{B}[I_{+,1}^{(A)}I_{+,1}^{(B)}(\alpha^{*})^{2}\gamma^{*}\sigma_{B}^{+}+I_{+,1}^{(A)}I_{-,1}^{(B)}(\alpha^{*})\gamma^{*}\sigma_{B}^{-}\\
&\nonumber +I_{+,1}^{(A)}I_{-,1}^{(B)*}(1+|\alpha|^{2})\gamma^{*}\sigma_{B}^{+}+I_{+,j}^{(A)}I_{-,j}^{(B)*}\gamma^{*}\sigma_{B}^{+}\\
&\nonumber +I_{+,1}^{(A)}I_{+,1}^{(B)*}(1+|\alpha|^{2})\gamma^{*}\sigma_{B}^{-}+I_{+,j}^{(A)}I_{+,j}^{(B)*}\gamma^{*}\sigma_{B}^{-}\\
&\nonumber +I_{-,1}^{(A)}I_{+,1}^{(B)}(\alpha^{*})^{2}\gamma\sigma_{B}^{+}+I_{-,1}^{(A)}I_{-,1}^{(B)}(\alpha^{*})^{2}\gamma\sigma_{B}^{-}\\
&\nonumber +I_{-,1}^{(A)}I_{-,1}^{(B)*}(1+|\alpha|^{2})\gamma\sigma_{B}^{+}+I_{-,j}^{(A)}I_{-,j}^{(B)*}\gamma\sigma_{B}^{+}\\
&\nonumber +I_{-,1}^{(A)}I_{+,1}^{(B)*}(1+|\alpha|^{2})\gamma\sigma_{B}^{-}+I_{-,j}^{(A)}I_{+,j}^{(B)*}\gamma\sigma_{B}^{-}\\
&\nonumber +I_{-,1}^{(A)*}I_{+,1}^{(B)}|\alpha|^{2}\gamma^{*}\sigma_{B}^{+}+I_{-,1}^{(A)*}I_{-,1}^{(B)}|\alpha|^{2}\gamma^{*}\sigma_{B}^{-}\\
&\nonumber +I_{-,1}^{(A)*}I_{-,1}^{(B)*}\alpha^{2}\gamma^{*}\sigma_{B}^{+}+I_{-,1}^{(A)*}I_{+,1}^{(B)*}\alpha^{2}\gamma^{*}\sigma_{B}^{-}\\
&\nonumber +I_{+,1}^{(A)*}I_{+,1}^{(B)}|\alpha|^{2}\gamma\sigma_{B}^{+}+I_{+,1}^{(A)*}I_{-,1}^{(A)}|\alpha|^{2}\gamma\sigma_{B}^{-}\\
&\nonumber +I_{+,1}^{(A)*}I_{-,1}^{(B)*}\alpha^{2}\gamma\sigma_{B}^{+}+I_{+,1}^{(A)*}I_{+,1}^{(B)*}|\alpha|^{2}\gamma\sigma_{B}^{-}]]\\
&\nonumber -\lambda_{B}\lambda_{A}[I_{+,1}^{(B)}I_{+,1}^{(A)}(\alpha^{*})^{2}\gamma^{*}\sigma_{B}^{+}+I_{+,1}^{(B)}I_{-,1}^{(A)}(\alpha^{*})^{2}\gamma\sigma_{B}^{+}\\
&\nonumber +I_{+,1}^{(B)}I_{-,1}^{(A)*}(1+|\alpha|^{2})\gamma^{*}\sigma_{B}^{+}+I_{+,j}^{(B)}I_{-,j}^{(A)*}\gamma^{*}\sigma_{B}^{+}\\
&\nonumber +I_{+,1}^{(B)}I_{+,1}^{(A)*}(1+|\alpha|^{2})\gamma\sigma_{B}^{+}+I_{+,j}^{(B)}I_{+,j}^{(A)*}\gamma\sigma_{B}^{+}\\
&\nonumber +I_{-,1}^{(B)}I_{+,1}^{(A)}(\alpha^{*})^{2}\gamma^{*}\sigma_{B}^{-}+I_{-,1}^{(B)}I_{-,1}^{(A)}(\alpha^{*})^{2}\gamma\sigma{B}^{-}\\
&\nonumber +I_{-,1}^{(B)}I_{-,1}^{(A)*}(1+|\alpha|^{2})\gamma^{*}\sigma_{B}^{-}+I_{-,j}^{(B)}I_{-,j}^{(A)*}\gamma^{*}\sigma_{B}^{-}\\
&\nonumber +I_{-,1}^{(B)}I_{+,1}^{(A)*}(1+|\alpha|^{2})\gamma\sigma_{B}^{-}+I_{-,j}^{(B)}I_{+,j}^{(A)*}\gamma\sigma_{B}^{-}\\
&\nonumber +I_{-,1}^{(B)*}I_{+,1}^{(A)}|\alpha|^{2}\gamma^{*}\sigma_{B}^{+}+I_{-,1}^{(B)*}I_{-,1}^{(A)}|\alpha|^{2}\gamma\sigma_{B}^{+}\\
&\nonumber +I_{-,1}^{(B)*}I_{-,1}^{(A)*}\alpha^{2}\gamma^{*}\sigma_{B}^{+}+I_{-,1}^{(B)*}I_{+,1}^{(A)*}\alpha^{2}\gamma\sigma_{B}^{+}\\
&\nonumber +I_{+,1}^{(B)*}I_{+,1}^{(A)}|\alpha|^{2}\gamma^{*}\sigma_{B}^{-}+I_{+,1}^{(B)*}I_{-,1}^{(A)}|\alpha|^{2}\gamma\sigma_{B}^{-}\\
&\nonumber +I_{+,1}^{(B)*}I_{-,1}^{(A)*}\alpha^{2}\gamma^{*}\sigma_{B}^{-}+I_{+,1}^{(B)*}I_{+,1}^{(A)*}(\alpha^{2})\gamma\sigma_{B}^{-}]\rho_{0,B}\\
&\nonumber -\lambda_{B}^{2}[(I_{+,1}^{(B)})^{2}(\alpha^{*})^{2}\sigma_{B}^{+}\rho_{0,B}\sigma_{B}^{+}+I_{+,1}^{(B)}I_{-,1}^{(B)}(\alpha^{*})^{2}\sigma_{B}^{+}\rho_{0,B}\sigma_{B}^{-}\\
&\nonumber +I_{+,1}^{(B)}I_{-,1}^{(B)*}(1+|\alpha|^{2})\sigma_{B}^{+}\rho_{0,B}\sigma_{B}^{+}+I_{+,j}^{(B)}I_{-,j}^{(B)*}\sigma_{B}^{+}\rho_{0,B}\sigma_{B}^{+}\\
&\nonumber +|I_{+,1}^{(B)}|^{2}(1+|\alpha|^{2})\sigma_{B}^{+}\rho_{0,B}\sigma_{B}^{-}+|I_{+,j}^{(B)}|^{2}\sigma_{B}^{+}\rho_{0,B}\sigma_{B}^{-}\\
&\nonumber +I_{-,1}^{(B)}I_{+,1}^{(B)}(\alpha^{*})^{2}\sigma_{B}^{-}\rho_{0,B}\sigma_{B}^{+}+(I_{-,1}^{(B)})^{2}(\alpha^{*})^{2}\sigma_{B}^{-}\rho_{0,B}\sigma_{B}^{-}\\
&\nonumber +|I_{-,1}^{(B)}|^{2}(1+|\alpha|^{2})\sigma_{B}^{-}\rho_{0,B}\sigma_{B}^{+}+|I_{-,j}^{(B)}|^{2}\sigma_{B}^{-}\rho_{0,B}\sigma_{B}^{+}\\
&\nonumber +I_{-,1}^{(B)}I_{+,1}^{(B)*}(1+|\alpha|^{2})\sigma_{B}^{-}\rho_{0,B}\sigma_{B}^{-}+I_{-,j}^{(B)}I_{+,j}^{(B)*}\sigma_{B}^{-}\rho_{0,B}\sigma_{B}^{-}\\
&\nonumber +I_{-,1}^{(B)*}I_{+,1}^{(B)}|\alpha|^{2}\sigma_{B}^{+}\rho_{0,B}\sigma_{B}^{+}+|I_{-,1}^{(B)}|^{2}|\alpha|^{2}\sigma_{B}^{+}\rho_{0,B}\sigma_{B}^{-}\\
&\nonumber +(I_{-,1}^{(B)*})^{2}\alpha^{2}\sigma_{B}^{+}\rho_{0,B}\sigma_{B}^{+}+I_{-,1}^{(B)*}I_{+,1}^{(B)*}\alpha^{2}\sigma_{B}^{+}\rho_{0,B}\sigma_{B}^{-}\\
&\nonumber +|I_{+,1}^{(B)}|^{2}|\alpha|^{2}\sigma_{B}^{-}\rho_{0,B}\sigma_{B}^{-}+I_{+,1}^{(B)}I_{-,1}^{(B)}|\alpha|^{2}\sigma_{B}^{-}\rho_{0,B}\sigma_{B}^{-}\\
&+ I_{+,1}^{(B)*}I_{-,1}^{(B)*}\alpha^{2}\sigma_{B}^{-}\rho_{0,B}\sigma_{B}^{+}+(I_{+,1}^{(B)})^{2}\alpha^{2}\sigma_{B}^{-}\rho_{0,B}\sigma_{B}^{-}].
\label{eq:secondorder2}
\end{align}
Now we can write the final quantum state of the target qubit in the cavity
\begin{align}
\nonumber\rho_{T,B}&=\rho_{0,B}+\rho_{T,B}^{(1)}+\rho_{T,B}^{(2)}\\
&\nonumber =\rho_{0,B}+\text{Tr}_{A}(\text{Tr}_{f}(U^{(1)}\rho_{0}))+\text{Tr}_{A}(\text{Tr}_{f}(U^{(1)}\rho_{0}))^{\dagger}\\
&\nonumber +\text{Tr}_{A}(\text{Tr}_{f}(U^{(2)}\rho_{0}))+
\text{Tr}_{A}(\text{Tr}_{f}(U^{(2)}\rho_{0}))^{\dagger}\\
&+\text{Tr}_{A}(\text{Tr}_{f}(U^{(1)}\rho_{0}U^{(1)\dagger}))
\end{align}
where we have now computed every term in this equation \eqref{eq:firstorder}, \eqref{eq:secondorder1}, and \eqref{eq:secondorder2}.

\section{APPENDIX: Additional plot}

This is a similar plot as the one in Figure 3 in the main text of the paper. For completeness, we show here the optimization of the Bloch sphere rotations to produce the maximum possible rotation in the $\eta$ direction. As advanced in the main text, the results are very much comparable to the $\Delta\phi$ maximization.

\begin{figure}[h]
\includegraphics[width=0.35\textwidth]{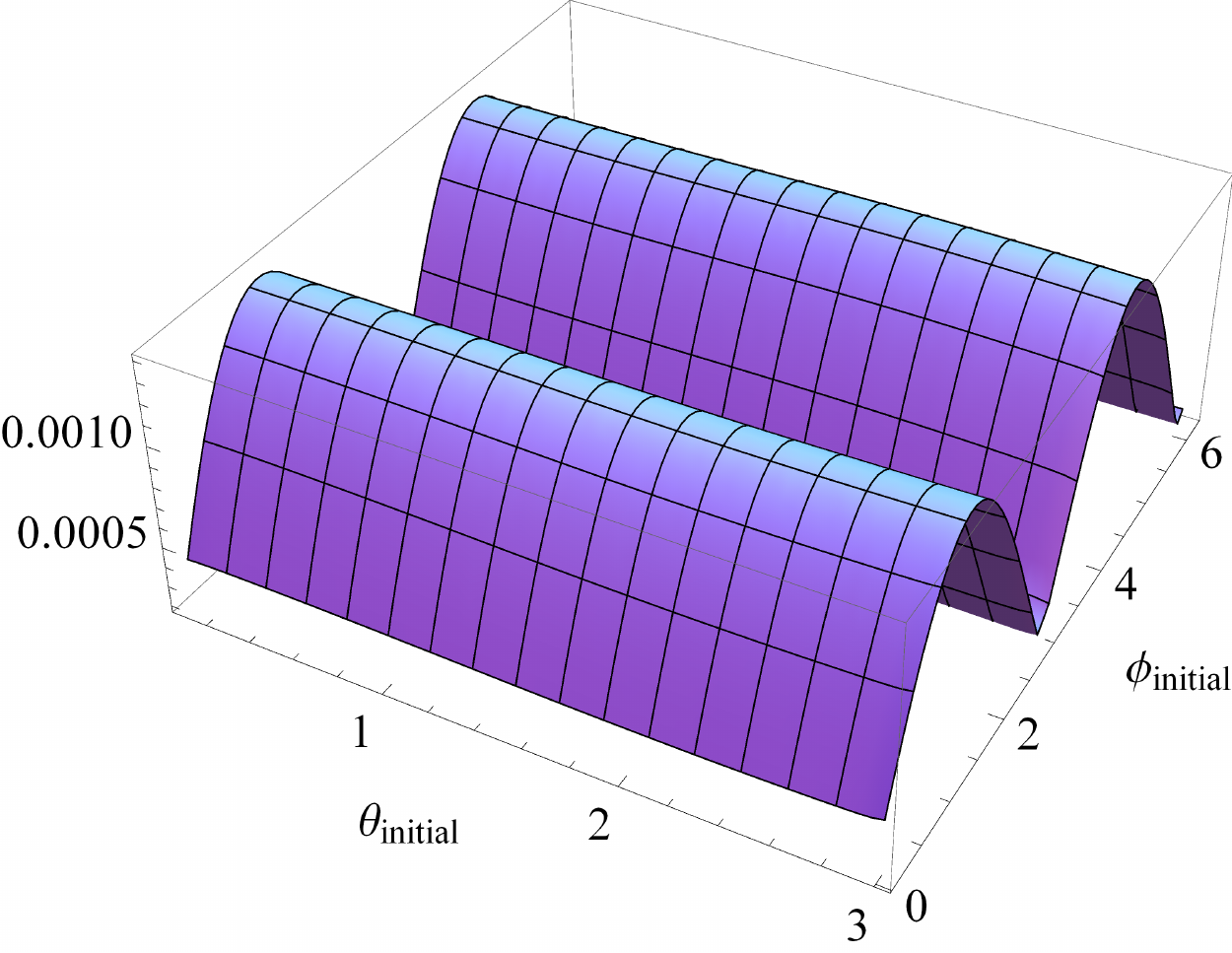}
\includegraphics[width=0.3\textwidth]{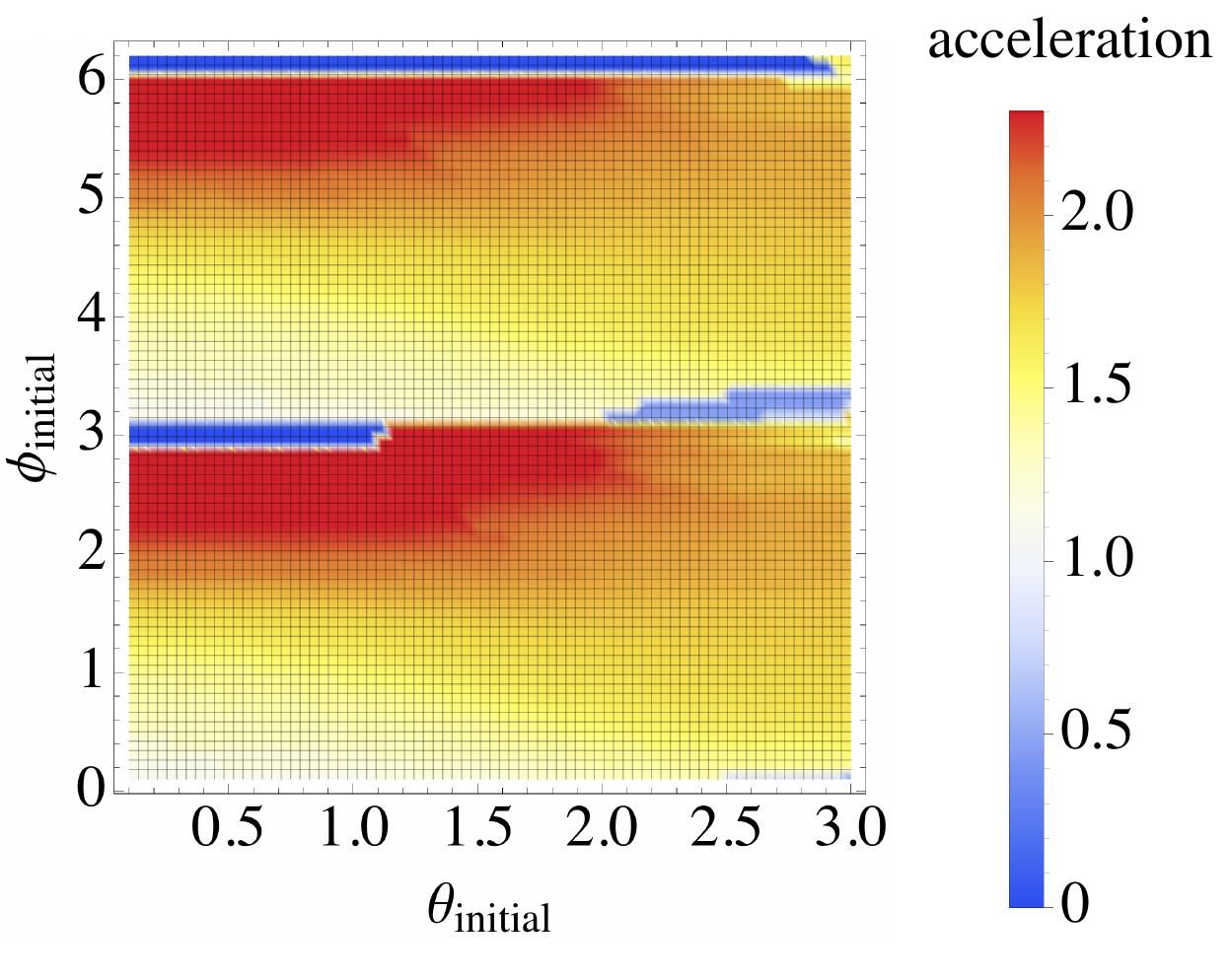}
\includegraphics[width=0.3\textwidth]{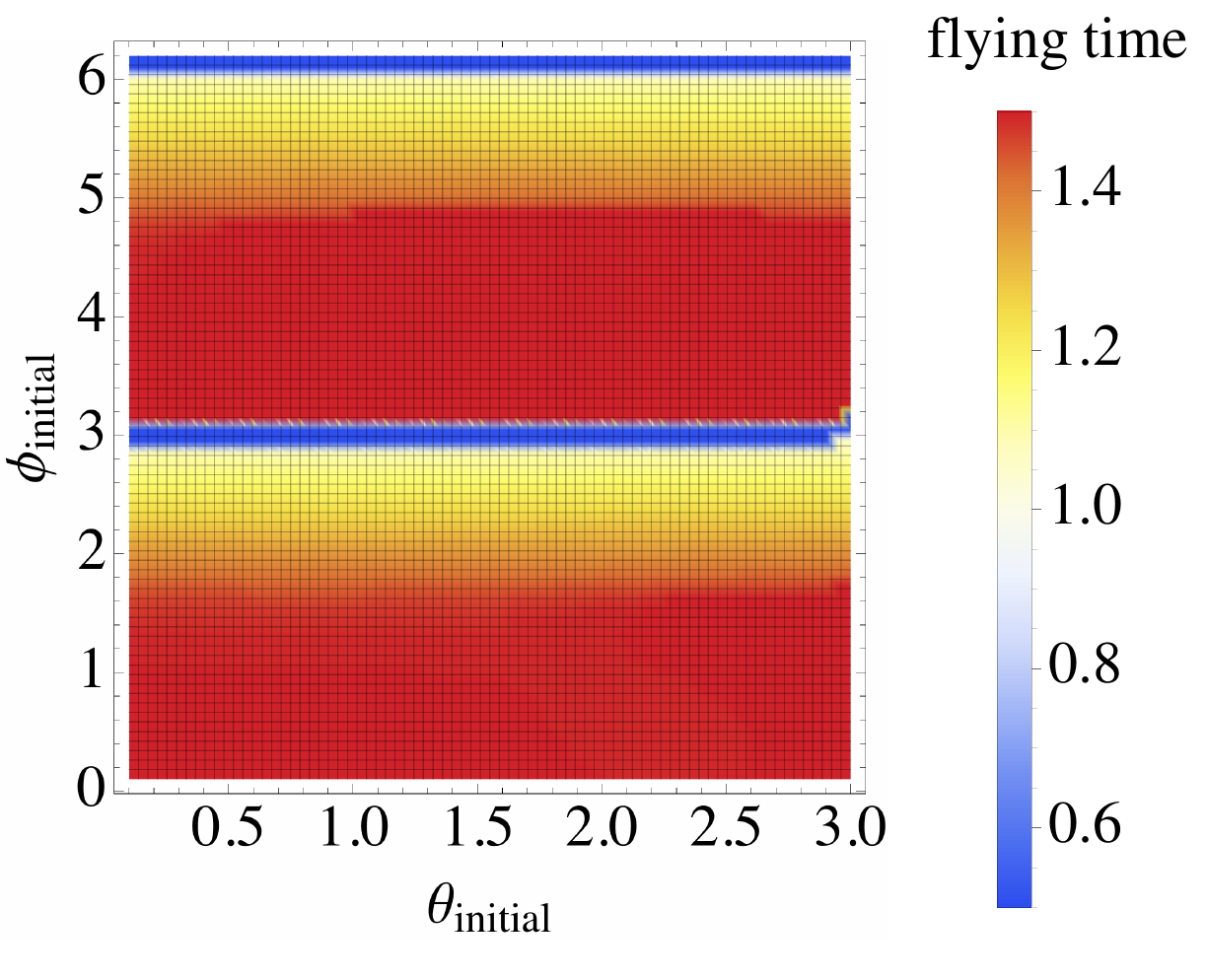}
  \caption{For $\lambda|\alpha|=0.01$ (where $|\alpha|$ is the amplitude of the coherent state), and for an initial probe state with $p=\frac{\ii}{\pi}$ we show (from top to bottom) {\bf a)} The change in the target qubit's azimuthal coordinate  $\Delta\theta$  maximized over $a\in[0,2.3]$ and $T\in[0,1.5]$ (natural units) for different initial pure states with all possible $\theta$ and $\phi$ on the Bloch sphere. {\bf b)} Accelerations of the probe atom which maximize $\Delta\theta$. In almost all regions, more relativistic accelerations optimize the magnitude of the rotation. {\bf c)} Flying times for the probe atom which maximize $\Delta\theta$. This shows again that when both the probe's acceleration and flying time are relativistic ($aT \sim c$), $\Delta\theta$ attains higher values. }
  \label{fig:unimportant}
\end{figure}

\bibliography{QCvRM}

\begin{thebibliography}{32}%
\makeatletter
\providecommand \@ifxundefined [1]{%
 \@ifx{#1\undefined}
}%
\providecommand \@ifnum [1]{%
 \ifnum #1\expandafter \@firstoftwo
 \else \expandafter \@secondoftwo
 \fi
}%
\providecommand \@ifx [1]{%
 \ifx #1\expandafter \@firstoftwo
 \else \expandafter \@secondoftwo
 \fi
}%
\providecommand \natexlab [1]{#1}%
\providecommand \enquote  [1]{``#1''}%
\providecommand \bibnamefont  [1]{#1}%
\providecommand \bibfnamefont [1]{#1}%
\providecommand \citenamefont [1]{#1}%
\providecommand \href@noop [0]{\@secondoftwo}%
\providecommand \href [0]{\begingroup \@sanitize@url \@href}%
\providecommand \@href[1]{\@@startlink{#1}\@@href}%
\providecommand \@@href[1]{\endgroup#1\@@endlink}%
\providecommand \@sanitize@url [0]{\catcode `\\12\catcode `\$12\catcode
  `\&12\catcode `\#12\catcode `\^12\catcode `\_12\catcode `\%12\relax}%
\providecommand \@@startlink[1]{}%
\providecommand \@@endlink[0]{}%
\providecommand \url  [0]{\begingroup\@sanitize@url \@url }%
\providecommand \@url [1]{\endgroup\@href {#1}{\urlprefix }}%
\providecommand \urlprefix  [0]{URL }%
\providecommand \Eprint [0]{\href }%
\providecommand \doibase [0]{http://dx.doi.org/}%
\providecommand \selectlanguage [0]{\@gobble}%
\providecommand \bibinfo  [0]{\@secondoftwo}%
\providecommand \bibfield  [0]{\@secondoftwo}%
\providecommand \translation [1]{[#1]}%
\providecommand \BibitemOpen [0]{}%
\providecommand \bibitemStop [0]{}%
\providecommand \bibitemNoStop [0]{.\EOS\space}%
\providecommand \EOS [0]{\spacefactor3000\relax}%
\providecommand \BibitemShut  [1]{\csname bibitem#1\endcsname}%
\let\auto@bib@innerbib\@empty
\bibitem [{\citenamefont {Montero}\ and\ \citenamefont
  {Mart\'in-Mart\'inez}(2011)}]{MigC}%
  \BibitemOpen
  \bibfield  {author} {\bibinfo {author} {\bibfnamefont {M.}~\bibnamefont
  {Montero}}\ and\ \bibinfo {author} {\bibfnamefont {E.}~\bibnamefont
  {Mart\'in-Mart\'inez}},\ }\href {\doibase 10.1007/JHEP07(2011)006} {\bibfield
   {journal} {\bibinfo  {journal} {J. High Energy Phys.}\ }\textbf {\bibinfo
  {volume} {2011}},\ \bibinfo {pages} {07} (\bibinfo {year}
  {2011})}\BibitemShut {NoStop}%
\bibitem [{\citenamefont {Mart\'{i}n-Mart\'{i}nez}\ \emph
  {et~al.}(2013{\natexlab{a}})\citenamefont {Mart\'{i}n-Mart\'{i}nez},
  \citenamefont {Aasen},\ and\ \citenamefont {Kempf}}]{aasen}%
  \BibitemOpen
  \bibfield  {author} {\bibinfo {author} {\bibfnamefont {E.}~\bibnamefont
  {Mart\'{i}n-Mart\'{i}nez}}, \bibinfo {author} {\bibfnamefont
  {D.}~\bibnamefont {Aasen}}, \ and\ \bibinfo {author} {\bibfnamefont
  {A.}~\bibnamefont {Kempf}},\ }\href {\doibase 10.1103/PhysRevLett.110.160501}
  {\bibfield  {journal} {\bibinfo  {journal} {Phys. Rev. Lett.}\ }\textbf
  {\bibinfo {volume} {110}},\ \bibinfo {pages} {160501} (\bibinfo {year}
  {2013}{\natexlab{a}})}\BibitemShut {NoStop}%
\bibitem [{\citenamefont {Bruschi}\ \emph {et~al.}(2013)\citenamefont
  {Bruschi}, \citenamefont {Dragan}, \citenamefont {Lee}, \citenamefont
  {Fuentes},\ and\ \citenamefont {Louko}}]{bruschi}%
  \BibitemOpen
  \bibfield  {author} {\bibinfo {author} {\bibfnamefont {D.~E.}\ \bibnamefont
  {Bruschi}}, \bibinfo {author} {\bibfnamefont {A.}~\bibnamefont {Dragan}},
  \bibinfo {author} {\bibfnamefont {A.~R.}\ \bibnamefont {Lee}}, \bibinfo
  {author} {\bibfnamefont {I.}~\bibnamefont {Fuentes}}, \ and\ \bibinfo
  {author} {\bibfnamefont {J.}~\bibnamefont {Louko}},\ }\href@noop {}
  {\bibfield  {journal} {\bibinfo  {journal} {Phy. Rev. Lett.}\ }\textbf
  {\bibinfo {volume} {111}},\ \bibinfo {pages} {090504} (\bibinfo {year}
  {2013})}\BibitemShut {NoStop}%
\bibitem [{\citenamefont {Friis}\ \emph {et~al.}(2012)\citenamefont {Friis},
  \citenamefont {Huber}, \citenamefont {Fuentes},\ and\ \citenamefont
  {Bruschi}}]{Prdvette}%
  \BibitemOpen
  \bibfield  {author} {\bibinfo {author} {\bibfnamefont {N.}~\bibnamefont
  {Friis}}, \bibinfo {author} {\bibfnamefont {M.}~\bibnamefont {Huber}},
  \bibinfo {author} {\bibfnamefont {I.}~\bibnamefont {Fuentes}}, \ and\
  \bibinfo {author} {\bibfnamefont {D.~E.}\ \bibnamefont {Bruschi}},\ }\href
  {\doibase 10.1103/PhysRevD.86.105003} {\bibfield  {journal} {\bibinfo
  {journal} {Phys. Rev. D}\ }\textbf {\bibinfo {volume} {86}},\ \bibinfo
  {pages} {105003} (\bibinfo {year} {2012})}\BibitemShut {NoStop}%
\bibitem [{\citenamefont {{Vandersypen}}\ \emph {et~al.}(2001)\citenamefont
  {{Vandersypen}}, \citenamefont {{Steffen}}, \citenamefont {{Breyta}},
  \citenamefont {{Yannoni}}, \citenamefont {{Sherwood}},\ and\ \citenamefont
  {{Chuang}}}]{nmr}%
  \BibitemOpen
  \bibfield  {author} {\bibinfo {author} {\bibfnamefont {L.~M.~K.}\
  \bibnamefont {{Vandersypen}}}, \bibinfo {author} {\bibfnamefont
  {M.}~\bibnamefont {{Steffen}}}, \bibinfo {author} {\bibfnamefont
  {G.}~\bibnamefont {{Breyta}}}, \bibinfo {author} {\bibfnamefont {C.~S.}\
  \bibnamefont {{Yannoni}}}, \bibinfo {author} {\bibfnamefont {M.~H.}\
  \bibnamefont {{Sherwood}}}, \ and\ \bibinfo {author} {\bibfnamefont {I.~L.}\
  \bibnamefont {{Chuang}}},\ }\href {\doibase 10.1038/414883a} {\bibfield
  {journal} {\bibinfo  {journal} {Nature}\ }\textbf {\bibinfo {volume} {414}},\
  \bibinfo {pages} {883} (\bibinfo {year} {2001})},\ \Eprint
  {http://arxiv.org/abs/quant-ph/0112176} {quant-ph/0112176} \BibitemShut
  {NoStop}%
\bibitem [{\citenamefont {Landulfo}\ and\ \citenamefont
  {Matsas}(2009)}]{matsako}%
  \BibitemOpen
  \bibfield  {author} {\bibinfo {author} {\bibfnamefont {A.~G.~S.}\
  \bibnamefont {Landulfo}}\ and\ \bibinfo {author} {\bibfnamefont {G.~E.~A.}\
  \bibnamefont {Matsas}},\ }\href@noop {} {\bibfield  {journal} {\bibinfo
  {journal} {Phys. Rev. A}\ }\textbf {\bibinfo {volume} {80}},\ \bibinfo
  {pages} {032315} (\bibinfo {year} {2009})}\BibitemShut {NoStop}%
\bibitem [{\citenamefont {C\'eleri}\ \emph {et~al.}(2010)\citenamefont
  {C\'eleri}, \citenamefont {Landulfo}, \citenamefont {Serra},\ and\
  \citenamefont {Matsas}}]{matsako2}%
  \BibitemOpen
  \bibfield  {author} {\bibinfo {author} {\bibfnamefont {L.~C.}\ \bibnamefont
  {C\'eleri}}, \bibinfo {author} {\bibfnamefont {A.~G.~S.}\ \bibnamefont
  {Landulfo}}, \bibinfo {author} {\bibfnamefont {R.~M.}\ \bibnamefont {Serra}},
  \ and\ \bibinfo {author} {\bibfnamefont {G.~E.~A.}\ \bibnamefont {Matsas}},\
  }\href {\doibase 10.1103/PhysRevA.81.062130} {\bibfield  {journal} {\bibinfo
  {journal} {Phys. Rev. A}\ }\textbf {\bibinfo {volume} {81}},\ \bibinfo
  {pages} {062130} (\bibinfo {year} {2010})}\BibitemShut {NoStop}%
\bibitem [{\citenamefont {DeWitt}(1967)}]{dewitt}%
  \BibitemOpen
  \bibfield  {author} {\bibinfo {author} {\bibfnamefont {B.~S.}\ \bibnamefont
  {DeWitt}},\ }\href@noop {} {\bibfield  {journal} {\bibinfo  {journal}
  {Physical Review}\ }\textbf {\bibinfo {volume} {160}},\ \bibinfo {pages}
  {1113} (\bibinfo {year} {1967})}\BibitemShut {NoStop}%
\bibitem [{\citenamefont {Crispino}\ \emph {et~al.}(2008)\citenamefont
  {Crispino}, \citenamefont {Higuchi},\ and\ \citenamefont
  {Matsas}}]{Crispino}%
  \BibitemOpen
  \bibfield  {author} {\bibinfo {author} {\bibfnamefont {L.~C.~B.}\
  \bibnamefont {Crispino}}, \bibinfo {author} {\bibfnamefont {A.}~\bibnamefont
  {Higuchi}}, \ and\ \bibinfo {author} {\bibfnamefont {G.~E.~A.}\ \bibnamefont
  {Matsas}},\ }\href@noop {} {\bibfield  {journal} {\bibinfo  {journal} {Rev.
  Mod. Phys.}\ }\textbf {\bibinfo {volume} {80}},\ \bibinfo {pages} {787}
  (\bibinfo {year} {2008})}\BibitemShut {NoStop}%
\bibitem [{\citenamefont {Scully}\ and\ \citenamefont
  {Zubairy}(1997)}]{ScullyBook}%
  \BibitemOpen
  \bibfield  {author} {\bibinfo {author} {\bibfnamefont {M.~O.}\ \bibnamefont
  {Scully}}\ and\ \bibinfo {author} {\bibfnamefont {M.~S.}\ \bibnamefont
  {Zubairy}},\ }\href@noop {} {\emph {\bibinfo {title} {Quantum Optics}}}\
  (\bibinfo  {publisher} {Cambridge University Press},\ \bibinfo {year}
  {1997})\BibitemShut {NoStop}%
\bibitem [{\citenamefont {Mart\'{i}n-Mart\'{i}nez}\ \emph
  {et~al.}(2013{\natexlab{b}})\citenamefont {Mart\'{i}n-Mart\'{i}nez},
  \citenamefont {Montero},\ and\ \citenamefont {del Rey}}]{eduardodewitt}%
  \BibitemOpen
  \bibfield  {author} {\bibinfo {author} {\bibfnamefont {E.}~\bibnamefont
  {Mart\'{i}n-Mart\'{i}nez}}, \bibinfo {author} {\bibfnamefont
  {M.}~\bibnamefont {Montero}}, \ and\ \bibinfo {author} {\bibfnamefont
  {M.}~\bibnamefont {del Rey}},\ }\href {\doibase 10.1103/PhysRevD.87.064038}
  {\bibfield  {journal} {\bibinfo  {journal} {Phys. Rev. D}\ }\textbf {\bibinfo
  {volume} {87}},\ \bibinfo {pages} {064038} (\bibinfo {year}
  {2013}{\natexlab{b}})}\BibitemShut {NoStop}%
\bibitem [{\citenamefont {Alhambra}\ \emph {et~al.}(2014)\citenamefont
  {Alhambra}, \citenamefont {Kempf},\ and\ \citenamefont
  {Mart\'in-Mart\'inez}}]{Alvaro}%
  \BibitemOpen
  \bibfield  {author} {\bibinfo {author} {\bibfnamefont {A.~M.}\ \bibnamefont
  {Alhambra}}, \bibinfo {author} {\bibfnamefont {A.}~\bibnamefont {Kempf}}, \
  and\ \bibinfo {author} {\bibfnamefont {E.}~\bibnamefont
  {Mart\'in-Mart\'inez}},\ }\href {\doibase 10.1103/PhysRevA.89.033835}
  {\bibfield  {journal} {\bibinfo  {journal} {Phys. Rev. A}\ }\textbf {\bibinfo
  {volume} {89}},\ \bibinfo {pages} {033835} (\bibinfo {year}
  {2014})}\BibitemShut {NoStop}%
\bibitem [{\citenamefont {Brown}\ \emph {et~al.}(2013)\citenamefont {Brown},
  \citenamefont {Mart\'{i}n-Mart\'{i}nez}, \citenamefont {Menicucci},\ and\
  \citenamefont {Mann}}]{brown}%
  \BibitemOpen
  \bibfield  {author} {\bibinfo {author} {\bibfnamefont {E.~G.}\ \bibnamefont
  {Brown}}, \bibinfo {author} {\bibfnamefont {E.}~\bibnamefont
  {Mart\'{i}n-Mart\'{i}nez}}, \bibinfo {author} {\bibfnamefont {N.~C.}\
  \bibnamefont {Menicucci}}, \ and\ \bibinfo {author} {\bibfnamefont {R.~B.}\
  \bibnamefont {Mann}},\ }\href {\doibase 10.1103/PhysRevD.87.084062}
  {\bibfield  {journal} {\bibinfo  {journal} {Phys. Rev. D}\ }\textbf {\bibinfo
  {volume} {87}},\ \bibinfo {pages} {084062} (\bibinfo {year}
  {2013})}\BibitemShut {NoStop}%
\bibitem [{\citenamefont {Takagi}(1986)}]{Takagi}%
  \BibitemOpen
  \bibfield  {author} {\bibinfo {author} {\bibfnamefont {S.}~\bibnamefont
  {Takagi}},\ }\href@noop {} {\bibfield  {journal} {\bibinfo  {journal} {Prog.
  Theor. Phys. Suppl.}\ }\textbf {\bibinfo {volume} {88}},\ \bibinfo {pages}
  {1} (\bibinfo {year} {1986})}\BibitemShut {NoStop}%
\bibitem [{\citenamefont {Reznik}\ \emph {et~al.}(2005)\citenamefont {Reznik},
  \citenamefont {Retzker},\ and\ \citenamefont {Silman}}]{Reznik}%
  \BibitemOpen
  \bibfield  {author} {\bibinfo {author} {\bibfnamefont {B.}~\bibnamefont
  {Reznik}}, \bibinfo {author} {\bibfnamefont {A.}~\bibnamefont {Retzker}}, \
  and\ \bibinfo {author} {\bibfnamefont {J.}~\bibnamefont {Silman}},\ }\href
  {http://link.aps.org/abstract/PRA/v71/e042104} {\bibfield  {journal}
  {\bibinfo  {journal} {Phys. Rev. A}\ }\textbf {\bibinfo {volume} {71}},\
  \bibinfo {eid} {042104} (\bibinfo {year} {2005})}\BibitemShut {NoStop}%
\bibitem [{\citenamefont {Mart\'{i}n-Mart\'{i}nez}\ \emph
  {et~al.}(2013{\natexlab{c}})\citenamefont {Mart\'{i}n-Mart\'{i}nez},
  \citenamefont {Brown}, \citenamefont {Donnelly},\ and\ \citenamefont
  {Kempf}}]{farming}%
  \BibitemOpen
  \bibfield  {author} {\bibinfo {author} {\bibfnamefont {E.}~\bibnamefont
  {Mart\'{i}n-Mart\'{i}nez}}, \bibinfo {author} {\bibfnamefont {E.~G.}\
  \bibnamefont {Brown}}, \bibinfo {author} {\bibfnamefont {W.}~\bibnamefont
  {Donnelly}}, \ and\ \bibinfo {author} {\bibfnamefont {A.}~\bibnamefont
  {Kempf}},\ }\href {\doibase 10.1103/PhysRevA.88.052310} {\bibfield  {journal}
  {\bibinfo  {journal} {Phys. Rev. A}\ }\textbf {\bibinfo {volume} {88}},\
  \bibinfo {pages} {052310} (\bibinfo {year} {2013}{\natexlab{c}})}\BibitemShut
  {NoStop}%
\bibitem [{\citenamefont {Mart\'{i}n-Mart\'{i}nez}\ \emph
  {et~al.}(2011)\citenamefont {Mart\'{i}n-Mart\'{i}nez}, \citenamefont
  {Fuentes},\ and\ \citenamefont {Mann}}]{BerryPh}%
  \BibitemOpen
  \bibfield  {author} {\bibinfo {author} {\bibfnamefont {E.}~\bibnamefont
  {Mart\'{i}n-Mart\'{i}nez}}, \bibinfo {author} {\bibfnamefont
  {I.}~\bibnamefont {Fuentes}}, \ and\ \bibinfo {author} {\bibfnamefont
  {R.~B.}\ \bibnamefont {Mann}},\ }\href {\doibase
  10.1103/PhysRevLett.107.131301} {\bibfield  {journal} {\bibinfo  {journal}
  {Phys. Rev. Lett.}\ }\textbf {\bibinfo {volume} {107}},\ \bibinfo {pages}
  {131301} (\bibinfo {year} {2011})}\BibitemShut {NoStop}%
\bibitem [{\citenamefont {Chen}\ and\ \citenamefont
  {Tajima}(1999)}]{Chen1999a}%
  \BibitemOpen
  \bibfield  {author} {\bibinfo {author} {\bibfnamefont {P.}~\bibnamefont
  {Chen}}\ and\ \bibinfo {author} {\bibfnamefont {T.}~\bibnamefont {Tajima}},\
  }\href {\doibase 10.1103/PhysRevLett.83.256} {\bibfield  {journal} {\bibinfo
  {journal} {Phys. Rev. Lett.}\ }\textbf {\bibinfo {volume} {83}},\ \bibinfo
  {pages} {256} (\bibinfo {year} {1999})}\BibitemShut {NoStop}%
\bibitem [{\citenamefont {Kazantsev}(1974)}]{ruso}%
  \BibitemOpen
  \bibfield  {author} {\bibinfo {author} {\bibfnamefont {A.~P.}\ \bibnamefont
  {Kazantsev}},\ }\href@noop {} {\bibfield  {journal} {\bibinfo  {journal} {Zh.
  Eksp. Teor. Fiz.}\ }\textbf {\bibinfo {volume} {66}},\ \bibinfo {pages}
  {1599} (\bibinfo {year} {1974})}\BibitemShut {NoStop}%
\bibitem [{\citenamefont {Arimondo}\ \emph {et~al.}(1977)\citenamefont
  {Arimondo}, \citenamefont {Inguscio},\ and\ \citenamefont
  {Violino}}]{hyperfine1}%
  \BibitemOpen
  \bibfield  {author} {\bibinfo {author} {\bibfnamefont {E.}~\bibnamefont
  {Arimondo}}, \bibinfo {author} {\bibfnamefont {M.}~\bibnamefont {Inguscio}},
  \ and\ \bibinfo {author} {\bibfnamefont {P.}~\bibnamefont {Violino}},\ }\href
  {\doibase 10.1103/RevModPhys.49.31} {\bibfield  {journal} {\bibinfo
  {journal} {Rev. Mod. Phys.}\ }\textbf {\bibinfo {volume} {49}},\ \bibinfo
  {pages} {31} (\bibinfo {year} {1977})}\BibitemShut {NoStop}%
\bibitem [{\citenamefont {et~al.}(2010)}]{Atlas1}%
  \BibitemOpen
  \bibfield  {author} {\bibinfo {author} {\bibfnamefont {G.~A.}\ \bibnamefont
  {et~al.}} (\bibinfo {collaboration} {ATLAS Collaboration}),\ }\href {\doibase
  10.1103/PhysRevLett.105.161801} {\bibfield  {journal} {\bibinfo  {journal}
  {Phys. Rev. Lett.}\ }\textbf {\bibinfo {volume} {105}},\ \bibinfo {pages}
  {161801} (\bibinfo {year} {2010})}\BibitemShut {NoStop}%
\bibitem [{\citenamefont {Leon}\ and\ \citenamefont
  {Sabin}(2008)}]{JleonSabin2a}%
  \BibitemOpen
  \bibfield  {author} {\bibinfo {author} {\bibfnamefont {J.}~\bibnamefont
  {Leon}}\ and\ \bibinfo {author} {\bibfnamefont {C.}~\bibnamefont {Sabin}},\
  }\href@noop {} {\bibfield  {journal} {\bibinfo  {journal} {Phys. Rev. A}\
  }\textbf {\bibinfo {volume} {78}},\ \bibinfo {pages} {052314} (\bibinfo
  {year} {2008})}\BibitemShut {NoStop}%
\bibitem [{\citenamefont {Leon}\ and\ \citenamefont
  {Sabin}(2009)}]{JleonSabin2b}%
  \BibitemOpen
  \bibfield  {author} {\bibinfo {author} {\bibfnamefont {J.}~\bibnamefont
  {Leon}}\ and\ \bibinfo {author} {\bibfnamefont {C.}~\bibnamefont {Sabin}},\
  }\href@noop {} {\bibfield  {journal} {\bibinfo  {journal} {Phys. Rev. A}\
  }\textbf {\bibinfo {volume} {79}},\ \bibinfo {pages} {012301} (\bibinfo
  {year} {2009})}\BibitemShut {NoStop}%
\bibitem [{\citenamefont {Sab\'in}\ \emph {et~al.}(2011)\citenamefont
  {Sab\'in}, \citenamefont {del Rey}, \citenamefont {Garc\'ia-Ripoll},\ and\
  \citenamefont {Le\'on}}]{resin}%
  \BibitemOpen
  \bibfield  {author} {\bibinfo {author} {\bibfnamefont {C.}~\bibnamefont
  {Sab\'in}}, \bibinfo {author} {\bibfnamefont {M.}~\bibnamefont {del Rey}},
  \bibinfo {author} {\bibfnamefont {J.~J.}\ \bibnamefont {Garc\'ia-Ripoll}}, \
  and\ \bibinfo {author} {\bibfnamefont {J.}~\bibnamefont {Le\'on}},\ }\href
  {\doibase 10.1103/PhysRevLett.107.150402} {\bibfield  {journal} {\bibinfo
  {journal} {Phys. Rev. Lett.}\ }\textbf {\bibinfo {volume} {107}},\ \bibinfo
  {pages} {150402} (\bibinfo {year} {2011})}\BibitemShut {NoStop}%
\bibitem [{\citenamefont {Nation}\ \emph {et~al.}(2009)\citenamefont {Nation},
  \citenamefont {Blencowe}, \citenamefont {Rimberg},\ and\ \citenamefont
  {Buks}}]{supercond}%
  \BibitemOpen
  \bibfield  {author} {\bibinfo {author} {\bibfnamefont {P.~D.}\ \bibnamefont
  {Nation}}, \bibinfo {author} {\bibfnamefont {M.~P.}\ \bibnamefont
  {Blencowe}}, \bibinfo {author} {\bibfnamefont {A.~J.}\ \bibnamefont
  {Rimberg}}, \ and\ \bibinfo {author} {\bibfnamefont {E.}~\bibnamefont
  {Buks}},\ }\href@noop {} {\bibfield  {journal} {\bibinfo  {journal} {Phys.
  Rev. Lett.}\ }\textbf {\bibinfo {volume} {103}},\ \bibinfo {pages} {087004}
  (\bibinfo {year} {2009})}\BibitemShut {NoStop}%
\bibitem [{\citenamefont {Wilson}\ \emph {et~al.}(2011)\citenamefont {Wilson},
  \citenamefont {Johansson}, \citenamefont {Pourkabirian}, \citenamefont
  {Simoen}, \citenamefont {Johansson}, \citenamefont {Duty}, \citenamefont
  {Nori},\ and\ \citenamefont {Delsing}}]{DCasimir}%
  \BibitemOpen
  \bibfield  {author} {\bibinfo {author} {\bibfnamefont {C.~M.}\ \bibnamefont
  {Wilson}}, \bibinfo {author} {\bibfnamefont {G.}~\bibnamefont {Johansson}},
  \bibinfo {author} {\bibfnamefont {A.}~\bibnamefont {Pourkabirian}}, \bibinfo
  {author} {\bibfnamefont {M.}~\bibnamefont {Simoen}}, \bibinfo {author}
  {\bibfnamefont {J.~R.}\ \bibnamefont {Johansson}}, \bibinfo {author}
  {\bibfnamefont {T.}~\bibnamefont {Duty}}, \bibinfo {author} {\bibfnamefont
  {F.}~\bibnamefont {Nori}}, \ and\ \bibinfo {author} {\bibfnamefont
  {P.}~\bibnamefont {Delsing}},\ }\href@noop {} {\bibfield  {journal} {\bibinfo
   {journal} {Nature}\ }\textbf {\bibinfo {volume} {479}},\ \bibinfo {pages}
  {376} (\bibinfo {year} {2011})}\BibitemShut {NoStop}%
\bibitem [{\citenamefont {Sab\'{i}n}\ \emph {et~al.}(2012)\citenamefont
  {Sab\'{i}n}, \citenamefont {Peropadre}, \citenamefont {del Rey},\ and\
  \citenamefont {Mart\'{i}n-Mart\'{i}nez}}]{PastFutPRL}%
  \BibitemOpen
  \bibfield  {author} {\bibinfo {author} {\bibfnamefont {C.}~\bibnamefont
  {Sab\'{i}n}}, \bibinfo {author} {\bibfnamefont {B.}~\bibnamefont
  {Peropadre}}, \bibinfo {author} {\bibfnamefont {M.}~\bibnamefont {del Rey}},
  \ and\ \bibinfo {author} {\bibfnamefont {E.}~\bibnamefont
  {Mart\'{i}n-Mart\'{i}nez}},\ }\href {\doibase 10.1103/PhysRevLett.109.033602}
  {\bibfield  {journal} {\bibinfo  {journal} {Phys. Rev. Lett.}\ }\textbf
  {\bibinfo {volume} {109}},\ \bibinfo {pages} {033602} (\bibinfo {year}
  {2012})}\BibitemShut {NoStop}%
\bibitem [{\citenamefont {del Rey}\ \emph {et~al.}(2012)\citenamefont {del
  Rey}, \citenamefont {Porras},\ and\ \citenamefont
  {Mart\'{i}n-Mart\'{i}nez}}]{Diegger}%
  \BibitemOpen
  \bibfield  {author} {\bibinfo {author} {\bibfnamefont {M.}~\bibnamefont {del
  Rey}}, \bibinfo {author} {\bibfnamefont {D.}~\bibnamefont {Porras}}, \ and\
  \bibinfo {author} {\bibfnamefont {E.}~\bibnamefont
  {Mart\'{i}n-Mart\'{i}nez}},\ }\href {\doibase 10.1103/PhysRevA.85.022511}
  {\bibfield  {journal} {\bibinfo  {journal} {Phys. Rev. A}\ }\textbf {\bibinfo
  {volume} {85}},\ \bibinfo {pages} {022511} (\bibinfo {year}
  {2012})}\BibitemShut {NoStop}%
\bibitem [{\citenamefont {Leibfried}\ \emph {et~al.}(2003)\citenamefont
  {Leibfried}, \citenamefont {Blatt}, \citenamefont {Monroe},\ and\
  \citenamefont {Wineland}}]{Leibfried2003}%
  \BibitemOpen
  \bibfield  {author} {\bibinfo {author} {\bibfnamefont {D.}~\bibnamefont
  {Leibfried}}, \bibinfo {author} {\bibfnamefont {R.}~\bibnamefont {Blatt}},
  \bibinfo {author} {\bibfnamefont {C.}~\bibnamefont {Monroe}}, \ and\ \bibinfo
  {author} {\bibfnamefont {D.}~\bibnamefont {Wineland}},\ }\href {\doibase
  10.1103/RevModPhys.75.281} {\bibfield  {journal} {\bibinfo  {journal} {Rev.
  Mod. Phys.}\ }\textbf {\bibinfo {volume} {75}},\ \bibinfo {pages} {281}
  (\bibinfo {year} {2003})}\BibitemShut {NoStop}%
\bibitem [{\citenamefont {Wallraff}\ \emph {et~al.}(2004)\citenamefont
  {Wallraff}, \citenamefont {Schuster}, \citenamefont {Blais}, \citenamefont
  {Frunzio}, \citenamefont {Huang}, \citenamefont {Majer}, \citenamefont
  {Kumar}, \citenamefont {Girvin},\ and\ \citenamefont
  {Schoelkopf}}]{Wallraff}%
  \BibitemOpen
  \bibfield  {author} {\bibinfo {author} {\bibfnamefont {A.}~\bibnamefont
  {Wallraff}}, \bibinfo {author} {\bibfnamefont {D.~I.}\ \bibnamefont
  {Schuster}}, \bibinfo {author} {\bibfnamefont {A.}~\bibnamefont {Blais}},
  \bibinfo {author} {\bibfnamefont {L.}~\bibnamefont {Frunzio}}, \bibinfo
  {author} {\bibfnamefont {R.-S.}\ \bibnamefont {Huang}}, \bibinfo {author}
  {\bibfnamefont {J.}~\bibnamefont {Majer}}, \bibinfo {author} {\bibfnamefont
  {S.}~\bibnamefont {Kumar}}, \bibinfo {author} {\bibfnamefont
  {S.}~\bibnamefont {Girvin}}, \ and\ \bibinfo {author} {\bibfnamefont
  {R.}~\bibnamefont {Schoelkopf}},\ }\href@noop {} {\bibfield  {journal}
  {\bibinfo  {journal} {Nature}\ }\textbf {\bibinfo {volume} {431}},\ \bibinfo
  {pages} {162} (\bibinfo {year} {2004})}\BibitemShut {NoStop}%
\bibitem [{\citenamefont {Niemczyk}\ \emph {et~al.}(2010)\citenamefont
  {Niemczyk}, \citenamefont {Deppe}, \citenamefont {Huebl}, \citenamefont
  {Menzel}, \citenamefont {Hocke}, \citenamefont {Schwarz}, \citenamefont
  {Garcia-Ripoll}, \citenamefont {Zueco}, \citenamefont {H{\"u}mmer},
  \citenamefont {Solano}, \citenamefont {Marx},\ and\ \citenamefont
  {Gross}}]{Solano}%
  \BibitemOpen
  \bibfield  {author} {\bibinfo {author} {\bibfnamefont {T.}~\bibnamefont
  {Niemczyk}}, \bibinfo {author} {\bibfnamefont {F.}~\bibnamefont {Deppe}},
  \bibinfo {author} {\bibfnamefont {H.}~\bibnamefont {Huebl}}, \bibinfo
  {author} {\bibfnamefont {E.~P.}\ \bibnamefont {Menzel}}, \bibinfo {author}
  {\bibfnamefont {F.}~\bibnamefont {Hocke}}, \bibinfo {author} {\bibfnamefont
  {M.~J.}\ \bibnamefont {Schwarz}}, \bibinfo {author} {\bibfnamefont {J.~J.}\
  \bibnamefont {Garcia-Ripoll}}, \bibinfo {author} {\bibfnamefont
  {D.}~\bibnamefont {Zueco}}, \bibinfo {author} {\bibfnamefont
  {T.}~\bibnamefont {H{\"u}mmer}}, \bibinfo {author} {\bibfnamefont
  {E.}~\bibnamefont {Solano}}, \bibinfo {author} {\bibfnamefont
  {A.}~\bibnamefont {Marx}}, \ and\ \bibinfo {author} {\bibfnamefont
  {R.}~\bibnamefont {Gross}},\ }\href@noop {} {\bibfield  {journal} {\bibinfo
  {journal} {Nature Physics}\ }\textbf {\bibinfo {volume} {6}},\ \bibinfo
  {pages} {772} (\bibinfo {year} {2010})}\BibitemShut {NoStop}%
\bibitem [{\citenamefont {Porras}\ and\ \citenamefont
  {Garc\'{i}a-Ripoll}(2012)}]{sidebands}%
  \BibitemOpen
  \bibfield  {author} {\bibinfo {author} {\bibfnamefont {D.}~\bibnamefont
  {Porras}}\ and\ \bibinfo {author} {\bibfnamefont {J.~J.}\ \bibnamefont
  {Garc\'{i}a-Ripoll}},\ }\href {\doibase 10.1103/PhysRevLett.108.043602}
  {\bibfield  {journal} {\bibinfo  {journal} {Phys. Rev. Lett.}\ }\textbf
  {\bibinfo {volume} {108}},\ \bibinfo {pages} {043602} (\bibinfo {year}
  {2012})}\BibitemShut {NoStop}%
\end{thebibliography}%

\end{document}